\documentclass[prb,preprintnumbers,amsmath,amssymb,floats]{revtex4}
\usepackage{graphicx}
\usepackage{dcolumn}
\usepackage{bm,epstopdf,soul}

\newcommand{\be}{\begin{equation}}
\newcommand{\ee}{\end{equation}}

\usepackage{amsmath}
\usepackage{color}
\usepackage{soul}

\begin{document}
\preprint{}

\title{Singular robust room-temperature spin response from topological Dirac fermions}


\author{Lukas Zhao,$^{1}$ Haiming Deng,$^1$ Inna Korzhovska,$^1$ Zhiyi Chen,$^1$  Marcin Konczykowski,$^2$
Andrzej Hruban,$^3$ Vadim Oganesyan,$^{4,5}$ \& Lia Krusin-Elbaum$^{1}$}

\vspace{3mm}

\small
\affiliation{$^1$Department of Physics, The City College of New York, CUNY, New York, NY 10031, USA \& The Graduate Center, CUNY, New York, NY 10016, USA} \affiliation{$^2$Laboratoire des Solides Irradi\'{e}s, CNRS UMR 7642 \& CEA-DSM-IRAMIS, Ecol\'{e} Polytechnique, F91128 Palaiseau cedex, France}
\affiliation{$^3$Institute of Electronic Materials Technology, 01-919 Warsaw, Poland}%
\affiliation{$^4$Department of Engineering Science and Physics, College of Staten Island, CUNY,  Staten Island, NY 10314, USA}
\affiliation{$^5$The Graduate Center, CUNY, New York, NY 10016, USA}%






\begin{abstract}
\noindent \small \textbf{
Topological insulators are a class of solids in which the nontrivial inverted bulk band structure gives rise to metallic surface states\cite{FuKanePRL07,HasanKane-RMPreview2010,Zhang-RMPreview2011,Chen-Science09,Hsieh-PRL09, Zhang-NatPhys09} that are robust against impurity scattering\cite{HasanKane-RMPreview2010,Zhang-RMPreview2011,Hsieh-Science09,HasanYazdani-Nature2009,Moore-PRL2010}. In three-dimensional (3D) topological insulators, however, the surface Dirac fermions intermix with the conducting bulk, thereby complicating access to the low energy (Dirac point) charge transport or magnetic response. Here we use differential magnetometry to probe spin rotation in the 3D topological material family (Bi$_2$Se$_3$, Bi$_2$Te$_3$, and Sb$_2$Te$_3$). We report a paramagnetic singularity in the magnetic susceptibility at low magnetic fields which persists up to room temperature, and which we demonstrate to arise from the surfaces of the samples. The singularity is universal to the entire family, largely independent of the bulk carrier density, and consistent with the existence of electronic states near the spin-degenerate Dirac point of the 2D helical metal. The exceptional thermal stability of the signal points to an intrinsic surface cooling process, likely of thermoelectric origin\cite{thermoelectricsNature2001,thermoelectric-GedikPRL2012}, and establishes a sustainable platform for the singular field-tunable Dirac spin response.
}

\end{abstract}

\maketitle
\normalsize
Enduring symbiosis between condensed matter physics and material science benefits whenever well established technological materials
turn out to be remarkably good model systems for fundamentally new physical phenomena which in turn can lead to disruptive technological advances.
Topological insulators are one recent example-- prized thermoelectrics \cite{thermoelectricsNature2001} since the 50's they also host topologically protected spin-helical surface states, as predicted by theory \cite{FuKanePRL07} (see Fig. 1a) and subsequently confirmed in a series of angularly resolved photoemission spectroscopy (ARPES) experiments \cite{Hsieh-Science09,HasanYazdani-Nature2009,Hsieh-Nature09}.  Much of the activity since has been inspired by prospects of harvesting exotic properties of these helical states for electrical manipulation of magnetic memory \cite{ME-Moore} and error-free topological quantum computing \cite{Majorana}.

Considerable effort is presently aimed at improving synthesis and characterization of these compounds with the goal of realizing materials with strongly suppressed bulk conduction channels -- the latter tend to obscure surface physics, a problem particularly severe in charge transport \cite{HasanKane-RMPreview2010,Zhang-RMPreview2011}. Indeed, complex intermixing (hybridization) of the bulk and surface states is clearly observed by a variety of surface probes; for example recent time-resolved ARPES experiments reveal strong phonon-assisted coupling between the surface and bulk electronic states at high lattice temperature and a unique cooling of Dirac fermions by acoustic phonons \cite{thermoelectric-GedikPRL2012}. `Aging' effects arising from complex surface reconstruction processes are also observed
\cite{surface-termination,Hofmann2DEG2012} -- they tend to promote formation of 2D electron gas states of bulk origin in close proximity to the topological Dirac surfaces.
Thus, existing materials continue to present a number of challenges
to complete understanding of the physics of topological Dirac metal, especially at low frequencies and on mesoscales. Magnetic susceptibility measurements reported in this work witness singular magnetic response of topological surface states, but also hint at an intriguing cooling process
involving these surface states and bulk carriers, thereby paving the way for systematic exploration of low energy electrodynamics of these transformative materials.

The experiments were performed using a weak low frequency \textit{ac} excitation field (see Fig. 1b and Methods) to probe the linear response, focussing on its in-phase component, which is the equilibrium susceptibility $\chi(B)=\partial M(B)/\partial H$ in the limit of zero frequency and  in a range of \textit{dc} fields $B=\mu_0 H$, including the vicinity of $B=0$ (see Supplementary Information, Section 1F).
Figure 1c shows susceptibility of the canonical $2^{nd}$ generation topological insulator Bi$_2$Se$_3$ measured in {\textit{dc} fields $H \parallel c$-axis (normal to the (00\={1}) cleavage surface)  of a platelike shaped crystal. Above $\sim 0.5~T$ the response is diamagnetic, consistent with a decades old magnetic susceptibility measurements \cite{old-diamag59}.
At lower fields, however, we detect a large cusplike \textit{paramagnetic} susceptibility that sharply rises above the diamagnetic `floor' in a narrow \textit{dc} field range of $\sim 0.2~\textrm{T}$ and approaching $\chi(H\rightarrow 0)$ in a straight line (Fig.~ 1c). This singularity arises from the sample's surface, is robust across all topological samples measured and is most naturally ascribed to the opening of a Zeeman gap \cite{Zhang-RMPreview2011} at the Dirac point of the helical metal. Before we turn to substantiating these claims we note one particularly spectacular aspect to our data -- its thermal stability. Indeed, the singular field dependence of the susceptibility shows no discernible signs of rounding up to the highest (room) temperature measured. This persistence of singular response to elevated temperature is remarkable and surprising when confronted with a rough conservative estimate of expected thermal smearing, \textit{e.g.} obtained from the ratio of thermal energy at $300~\textrm{K}$ ($ \simeq 27~\textrm{meV}$) to the rather small bulk gap of these materials $\sim 100-300~\textrm{meV}$.

The presence of the cusp in near-zero-field susceptibility is universal -- it is observed in all three topological insulators: Sb$_2$Te$_3$,  Bi$_2$Te$_3$, and Bi$_2$Se$_3$ (Fig.~2a-2c). It is absent in all our calibration and background materials (see Supplementary Information Section C1, Fig. S1),
which were carefully screened for any spurious signals. At higher fields, $H \gtrsim 0. 5~T$, the temperature-dependent diamagnetism dominates (Fig.~2d-2f, and Fig. S2);
it appears to correlate with the details of the bulk band structure, but less clearly with the particulars of donor (\textit{n}-type) or acceptor (\textit{p}-type) intrinsic defects (Fig.~2g-2i) present in the bulk.

The height of the cusp is evidently sensitive somewhat to the density of defects quenched in during the crystal growth (see Methods), and there is an aging effect  \cite{Hofmann2DEG2012} that can reduce the height over time by an appreciable (up to 5) factor (an example is shown in Fig. S3).
The absolute magnitude of the cusp in different crystals varies some with the intrinsic bulk carrier density, which in any particular crystal is determined from  the measurements of Hall conductivity (see Fig.~2) or Shubnikov-de-Haas (SdH) quantum oscillations (Fig. S4).
However the `cuspiness' as quantified by the $B = \mu_0 H \rightarrow 0$ slope for any given member of this topological insulator family is universal. An example of this is shown in Fig. 3a, where we compare two Bi$_2$Te$_3$ crystals with carrier concentrations differing by two orders of magnitude. The cusp is frequency independent (Fig.~4a and Fig. S6),
as expected for such low frequency response ($2\sim10$ kHz).
This is further confirmed by the singular signal in the differential susceptibility obtained from the \textit{dc} magnetization measurements using Superconducting Quantum Interference Device (SQUID) magnetometer, see Fig. S6d.
And finally, the `smoking-gun' evidence that the cusp originates from the surface states is illustrated in Fig.~3b, which shows that for the same crystal area, when the sample thickness is reduced the height of the cusp remains unchanged, while the diamagnetic background closely scales with the volume. We note that similar, albeit weaker, response is detected with the sample rotated by $90~\textrm{degrees}$ (see Fig. S7),
consistent with the signal originating from noncleaving surfaces \cite{other-surfaces-2012} where the Dirac dispersion is more complex.

Our finding of a prominent singular magnetic response that survives high temperatures, huge variations in carrier density, and does not scale with sample volume is quite surprising and as far as we know unprecedented. Absent any paramagnetic impurities (see Methods) or signs of itinerant ferromagnetism, the origin of this particular low field anomaly may be traced most naturally to the ungapped Dirac point. The simplest description of the Dirac fermions is captured by a non-interacting Rashba-type \cite{Rashba} Hamiltonian that effectively locks electron spin to its momentum, \textit{i.e.} parallel to the sample's surface (see Supplemental Information, Sec. 2). The effect of magnetic field applied transverse to the surface enters through a Zeeman coupling which we treat explicitly and via orbital quantization which we ignore (this approximation is justified by the absence of oscillatory effects at low fields in our experiments, and, \emph{a posteriori}, we can also confirm that Dirac Landau level spacing is essentially negligible compared to the Zeeman gap in the parameter range relevant to our experiments -- see Supplementary Information, Section 2).
The equilibrium susceptibility is obtained by taking the $2^{nd}$ derivative  of the total free energy with respect to magnetic field $B$.
With both chemical potential $\mu$ and temperature set to zero, low field \emph{areal} (sheet) susceptibility $\chi_A$ (see Supplementary Information) reduces to
\begin{equation}
\chi_A(B) \cong \frac{\mu_0}{4 \pi^2}\left[\frac{(g \mu_B)^2 \Lambda }{\hbar  v_F}-\frac{2 (g \mu_B)^3}{\hbar ^2 v_F^2}|B|+\ldots\right],
\end{equation}
where $g$ is the Land\'{e} $g$-factor and $v_F$ is the Fermi velocity. This paramagnetic Dirac susceptibility has the form of a cusp with a linear-in-field decay at low fields, just as the cusp observed in our experiments (Fig. ~3c).  The maximum of $\chi_A$ depends on the effective size of the momentum space $\Lambda$ contributing to the singular part of the free energy, and thus may be controlled in part by hexagonal warping of the Dirac cone \cite{HexWarpFu2009} and by the details of the bulk bands. However, the singular \emph{field dependence} only depends on universal (low energy) parameters  through the slope $\frac {2(g \mu_B)^3}{\hbar ^2 v_F^2}$ of $\chi$ in the limit $ B \rightarrow 0$. To compare with the experiment we write the total susceptibility as a sum of the background contribution $\chi_0$ and surface contribution $\chi=\chi_0+\chi_A x/L_z$, where $x$ is the fraction of the surface contributing and $L_z\approx 1\ mm$ is sample's thickness. We obtain a good match to the shape and the magnitude of the cusp (see Fig.~3c) by using parameter values consistent with the reported velocity $v_F$ in Bi$_2$Te$_3$ from Landau level spectroscopy \cite{Wolos2012} and large effective \textit{g}-factor \cite{Boebinger-g-factor}, broadly consistent with the overall scale of \textit{g}-factors expected for topological insulators and obtained from our SdH measurements (Figs.~3c and S4).
The participating surface fraction that emerges from this analysis is remarkably small, $x\approx 0.002$, \textit{i.e.} these states are very rare.

The existence of the sharp nonanalytic paramagnetic cusp at zero temperature requires the surface Fermi level to be at the Dirac point, $\mu=0$. Otherwise, for $\mu\neq 0$, we expect a smooth dependence (rounding)  near $B=0$ with sharp jump singularities in $\chi$ on a field scale $\delta B = \mu/(g \mu_B)$ where the Fermi level enters the valence or conduction band.
Further phenomenological description can be facilitated by recasting the low field paramagnetic response in Eq.~1 in terms of effective Dirac bandwidth $W = \hbar v_F \Lambda$ and field energy $E_B=g \mu_B B$ as $\chi_A(B)=\frac{(g \mu_B)^2 \Lambda^2 }{W}\left ( 1-\frac{2 E_B}{W}+\ldots\right)$, so
the characteristic width of the cusp is set by the condition $W \approx E_B$. The observed temperature insensitivity requires that the thermal energy $E_T = k_B T \ll E_B<W$, or $T\lesssim 10K$, which may be relaxed somewhat on the level of this simple phenomenology if both \textit{g}-factor and Fermi velocity are temperature dependent (Supplementary Information, Section 2).

In our \emph{ac} experiments, no appreciable rounding of the cusp is observed --  this finding is profoundly unexpected in view of 
the location of Fermi level gleaned from ARPES or STM. Separate experimental work will be required to obtain a clear and detailed understanding of the microscopic origins of the electronic states \cite{Franz-PRB2012} giving rise to the singular response. From the established surface nature and the observed aging effects we infer that renormalization of the effective potential near the sample's surface in the course of aging is important. Also, the remarkable robustness to the variation in bulk carrier density and therefore bulk screening length, suggests that electrostatic models invoking bulk dopants as the dominant source of disorder at the surface may not be adequate to capture these states. Such models do readily produce large scale inhomogeneities of chemical potential, $\mu$, which have been observed, for example, in graphene \cite{Martin-graphene-puddles} and has been recently directly mapped in several topological insulators via scanning tunneling microscopy (STM) \cite{Beidenkopf-NatPhys2010}. 
The typical amplitude of inhomogeneity in the latter study, $10\sim 20$ meV, appears too small to couple to the electronic states near the Dirac point. However, rare states, that based on our analysis occupy only $\approx 0.2\%$ of sample's surface, may not be readily observed in STM.
Moreover, the role played by unavoidable differences in surface preparation among different experiments remains to be established.

Yet another intriguing finding in our experiments is the apparent thermal stability of the singular \emph{ac} response. 
This is certainly not within our simple Dirac phenomenology, which has in it scales on the order of only $10$ K. In fact, we may argue that \emph{any} equilibrium theory of the singular response in these narrowband semiconductors must show thermal effects near room temperature, as the band gap is only a few times larger, at best. Indeed, in \textit{dc} magnetization measurements using SQUID the singular response at higher temperatures is rounded (Fig. S6d). We propose, therefore, that the local temperature at the location of electronic states responsible for the cusp is, in fact, strongly affected by the \emph{ac} probe itself, \textit{i.e.} these patches are kept at very low, possibly cryogenic effective temperature even though the cryostat and the rest of the sample are "warm".
One plausible, albeit still  speculative, scenario (see Fig. 4) for this invokes disorder as the origin of local Peltier elements.  The most natural source of power for the putative Peltier cooler is the rather large eddy current which does not contribute to $\chi$ itself but rather to the imaginary, out-of-phase part of $\chi(\omega)$ (Fig.~S5).
To suppress Peltier heating (unavoidable due to \emph{ac} excitation), this would require a rectifying element as well (see Fig. 4c and Fig. S9).
From general consideration of the rectification process there should be then second harmonic generation, which we clearly observe (Fig. 4b and Fig. S10).
The above scenario implies strong enhancement of the effective (local) thermoelectric figure of merit as compared to known bulk values for these materials (see, \textit{e.g.,} Ref. \onlinecite{reduced_surface_e-ph_2013}), which would be natural, based on the existing work on improved thermoelectricity in nano-constrictions \cite{nano-constrictions,spin-rectifier}, and on strong frequency dependence of the transport coefficients under geometric confinement, as in the case of phonon heat conductivity \cite{frequency-dependent-thermal}. We also note that strong (local) variations of material properties, \textit{e.g.} due to the presence of disorder, can give rise to a novel variant of thermoelectric cooling, a ``Thompson cooler", which has been predicted to display significant improvement of performance and,  in principle, enable cooling to very low, even cryogenic temperatures \cite{Thomson-cooler}. Detailed theory of the mechanism of thermal stability is beyond the scope of this work and should be further explored.

Our experiments document a singularity in the low field response in a whole family of materials with topological surface states which does not arise from either strong correlations or fine tuning the chemical potential to the Dirac point. They are profoundly counterintuitive as they suggest the controlling role of rare states (patches) near the Dirac point realized under generic surface conditions in these samples.  With this assumption we are able to reproduce the overall shape and magnitude of the response. One of the surprising quantitative insights that emerged was that a minority ($\approx 0.2\%$) of the surface is responsible for the singular signal. This simple  phenomenology is a step forward to a
precise theoretical understanding and improved experimental control of these phenomena that will be crucial for manipulating robust polarization of protected surface states at room temperature.

\vspace{2mm}

\noindent {\textbf{Methods}

\small \noindent Single crystals of Bi$_2$Se$_3$, Bi$_2$Te$_3$, and Sb$_2$Te$_3$ were grown by a modified Bridgman method (using evacuated quartz tubes in a horizontal gradient furnace heated to 1000$^o$ C and cooled to room temperature in 7 days)  or the standard Bridgman-Stockbarger method \cite{Wolos2012} using a vertical temperature gradient pull. The starting materials used in modified Bridgman were cm-sized chunks of Sb, Bi (purity of both 99.9999\%), Te (purity 99.9995\%), and Se (99.995\%) from Alfa-Aesar used in stoichiometric ratios. X-ray diffraction of crystals was performed in Panalytical diffractometer using Cu K$\alpha~ (\lambda = 1.5405{\AA})$ line from Philips high intensity ceramic sealed tube (3~kW) X-ray source with a Soller slit (0.04 rad) incident and diffracted beam optics. The impurity level determined by elemental analysis using glow discharge mass spectrometry was found to be less than 0.005 ppm wt. We used a series of crystals with different carrier densities (set by the number of charged vacancies and antisites quenched in during the crystal growth) which were obtained by varying the speed (down to 2~mm/hr) of the pull or the gradient profile in a horizontal or vertical setup. Carrier densities were determined from the measurements of Hall resistivity and Shubnikov-de Haas oscillations (see Supplementary Information). All crystals were exfoliated to expose fresh surfaces prior to measurements, with the exception of surface `aging' studies.  Differential susceptibility measurements were performed in a Quantum Design PPMS system, in a compensated pickup-coil detection configuration (Fig.~1b) with the \textit{ac} excitation and detection coils designed to align with the the direction of applied static field. The \textit{ac} excitation field amplitude was set at $10^{-5}~\textrm{T}$ in a frequency range up to 10 kHz. Measurements of the sample holder, starting materials, NbSe$_2$, and furnace annealed Te were performed to exclude any possible contamination and systemic contributions (Supplementary Information).
The system was calibrated using paramagnetic Pd standard, see Fig.~S1d. The field scans at different temperatures over a larger field range for the topological insulators in this study are shown in the Supplementary Information. Calculations were performed using Mathematica.

\noindent \textbf{Acknowledgements} We greatly appreciate the insights of Kyunghwa Park and thank Gil Refael for his useful suggestions and comments. We gratefully acknowledge Glen Kowach for his generous help and expert advice with the Bridgman crystal growth and Agnieszka Wo{\l}o\'{s} for selecting crystals with low carrier density. This work was supported by the NSF DMR-1122594 and DOD-W911NF-13-1-0159 (L.K.-E.), and DMR-0955714 (V.O.).

\vspace{1mm}

\noindent \textbf{Author contributions} Experiments were designed by L.Z. and L.K.-E..  L.Z. and H.D. carried out the growth of single crystals, M.K. and A.H. provided Bi$_2$Te$_3$ crystals with the lowest carrier densities, and I.K. and Z.C. performed structural and chemical characterization of all crystals. \textit{ac} susceptibility measurements were done by L.Z. and H.D., data analysis was done by L.Z. and L.K.-E. Dirac phenomenology and the mechanism of Peltier cooling were formulated jointly by V.O. and L.K.-E. L.K.-E. and V.O. wrote the manuscript with critical input from L.Z.

\vspace{1mm}

\noindent \textbf{Additional information} The authors declare that they have no competing financial interests. Supplementary information accompanies this paper on www.nature.com/naturematerials. Correspondence and requests for materials should be addressed to L. K.-E.

\normalsize

\normalsize

\newpage%

\noindent\section*{FIGURE LEGENDS}

\noindent \textbf{Figure 1 $\mid$ Dirac point origin of the large singular spin susceptibility near zero magnetic field.}
\textbf{a}, The energy-momentum relation of the surface states in a 3D topological insulator has a spin-helical Dirac cone structure arising from strong spin-orbit interaction that locks spins to their momentum \cite{Hsieh-Nature09}. For the (00\={1}) surfaces parallel to the quintuple layers \cite{HasanKane-RMPreview2010} of a layered topological insulator such as Bi$_2$Se$_3$ the spin texture near the Dirac point is riding on a circular constant energy contours of the Dirac bands, with spins aligned along normal to the momentum. At the Dirac point, however, electron spins should be free to align along the tiny field as long as the Dirac spectrum is not gapped. \textbf{b}, Magnetic susceptibility of Bi$_2$Se$_3$ measured by applying a small \textit{ac} excitation field $h_{ac}$ (see Methods) shows that \textbf{c}, spin response  is cusp-like and large near zero applied \textit{dc} magnetic field.
The susceptibility cusp is remarkably robust up to room temperature for both, $H \parallel c$-axis and $H \parallel ab$ field directions, see Fig.~S7. It rides on a temperature dependent diamagnetic background, see Fig.~S2. Here, the data at different temperatures were shifted to the lowest temperature of this study  to indicate that both the slope and the height of the cusp between $1.9~\textrm{K}$ and $300~\textrm{K}$ remain intact.

\vspace{5mm}

\noindent \textbf{Figure 2 $\mid$ Universality of singular spin response near zero magnetic field.} The zero-field susceptibility cusp is found in all three topological insulators: \textbf{a}, Sb$_2$Te$_3$, \textbf{b}, Bi$_2$Te$_3$, and \textbf{c}, Bi$_2$Se$_3$. The susceptibility surface in the $H-T$ phase space for fields above $H \sim 0. 5~T$ is shown  in \textbf{d}, for Sb$_2$Te$_3$, in \textbf{e}, for Bi$_2$Te$_3$, and in \textbf{f}, for Bi$_2$Se$_3$ (see Supplementary Information). The most pronounced temperature dependence is found in Sb$_2$Te$_3$ (\textbf{d}), which has the smallest bulk bandgap of $\sim 100~\textrm{meV}$. \textbf{g}-\textbf{i}, Corresponding schematic band structures \cite{Zhang-NatPhys09} indicate noticeable differences in the location of the Dirac point relative to the bulk valence and conduction bands. Measurements of Hall resistivity (\textbf{g}-\textbf{i}) show  that Te-based TIs, Sb$_2$Te$_3$ and Bi$_2$Te$_3$, are intrinsically \textit{p}-type, while the Se-based TI,  Bi$_2$Se$_3$ is \textit{n}-type.

\vspace{5mm}

\noindent \textbf{Figure 3 $\mid$ Signatures of the surface origin of the cusp.}
\textbf{a}, Susceptibility cusp for two Bi$_2$Te$_3$ crystals with carrier densities differing by two orders of magnitude. The slope of the cusp is independent of the bulk carrier density $n$. Here the diamagnetic background was subtracted and the height of the cusp was normalized to $\chi (B =0)$, which for the $n \sim  10^{19}~\textrm{cm}^{-3}$ crystal was $3 \times 10^{-5}~ \textrm{emu/cc}$, and for the $n \sim  10^{17}~\textrm{cm}^{-3}$ crystal was $3.5 \times 10^{-5}~ \textrm{emu/cc}$. \textbf{b}, Left: Susceptibility cusp before and after cutting the crystal thickness by a factor of 0.63 (red), 0.29 (green), and 0.15 (blue) appears to be independent of thickness $t$. The diamagnetic background scales with thickness (volume for the fixed sample area $A$). Right: The data for all thicknesses shown on the left shifted to match the diamagnetic background. The signal to noise decreases with sample volume.  \textbf{c}, The simple Dirac model of Eq. 1 produces a very good match to the data, as illustrated for the case of  Sb$_2$Te$_3$ (see also Supplemental Information). Here $\chi=x \chi_A/L_z$ and $\chi_A$ is the 2D susceptibility of the Dirac state, $L_z\approx 10^{-3}m$, thickness of our samples, and $x<1$ the effective areal fraction occupied by the ungapped Dirac state ($x$ is used as a fitting parameter). Other parameter values used to generate this plot are $\mu=k_B T=0, g=60, v_F=2\cdot 10^3 m/s$, which are known from our own studies (see Supplemental Information) and those of others \cite{Wolos2012}. Both $x$ and $\Lambda$ (effective radius of {\bf k}-space contributing to singular response) were adjusted to match the data, producing $x\approx 0.002$ and $\Lambda=5\cdot 10^8 m^{-1}$. The cusp is preserved even when hexagonal warping (inset in \textbf{c}) is taken into account \cite{HexWarpFu2009} -- it is merely subsumed into $\Lambda$.  \textbf{d},
Rare regions of chemical potential $\mu \approx 0$ (grey) can exist in-between electron (blue) and hole (yellow) droplets due in part to electrostatic potential established by the charged defects in the bulk \cite{Beidenkopf-NatPhys2010}. Such fluctuations of the local surface charge are likely ``healing" in the course of the aging process \cite{surface-termination,Hofmann2DEG2012} as the mean chemical potential steadily floats \emph{away} from the Dirac point towards bulk conduction or valence bands, as has been documented in ARPES studies \cite{surface-termination,Hofmann2DEG2012}. This is qualitatively consistent with the observed decrease in the amplitude of the paramagnetic anomaly over time.

\vspace{5mm}

\noindent \textbf{Figure 4 $\mid$ Surface cooling by the bulk.}
\textbf{a}, The in-phase component of the susceptibility containing the singular cusp is frequency independent (shown here for Sb$_2$Te$_3$). However, the \textit{diamagnetic} susceptibility is slightly frequency dependent (see Fig. S6). \textbf{b}, The nonlinearity of the surface-bulk connection is witnessed by the observed $2^{nd}$ harmonic of $\chi$. It is consistent with the existence of "rectifying" paths in the putative thermoelectric cooling elements, see Figs. S9 and S10 and discussion in Supplementary Information) required for the cooling of small fraction of sample's surface and thus suppressing thermalization of Dirac surfaces with the bulk, as explained in text. The effective cooling of the surface is naturally achieved by the electron and hole puddles in the sub-surface region forming a Peltier element (inset) owing its cooling efficiency partly to nanoconstriction and partly to frequency-dependent transport coefficients \cite{frequency-dependent-thermal}.

\newpage

\newpage
\eject

\begin{center}

\includegraphics[width=16cm]{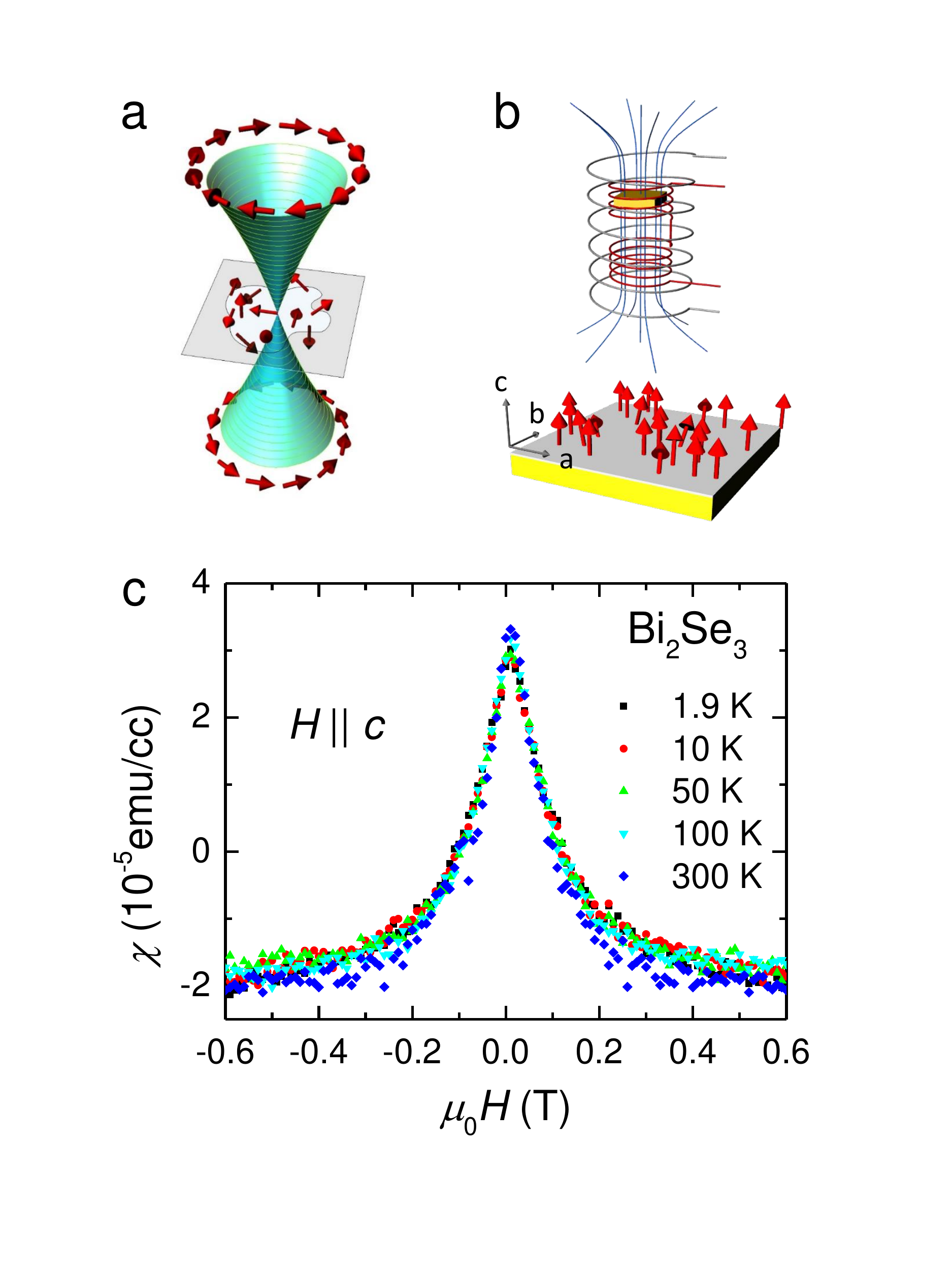}
\vfill\hfill Fig.~1 Zhao {\it et al.} \eject

\includegraphics[width=18cm]{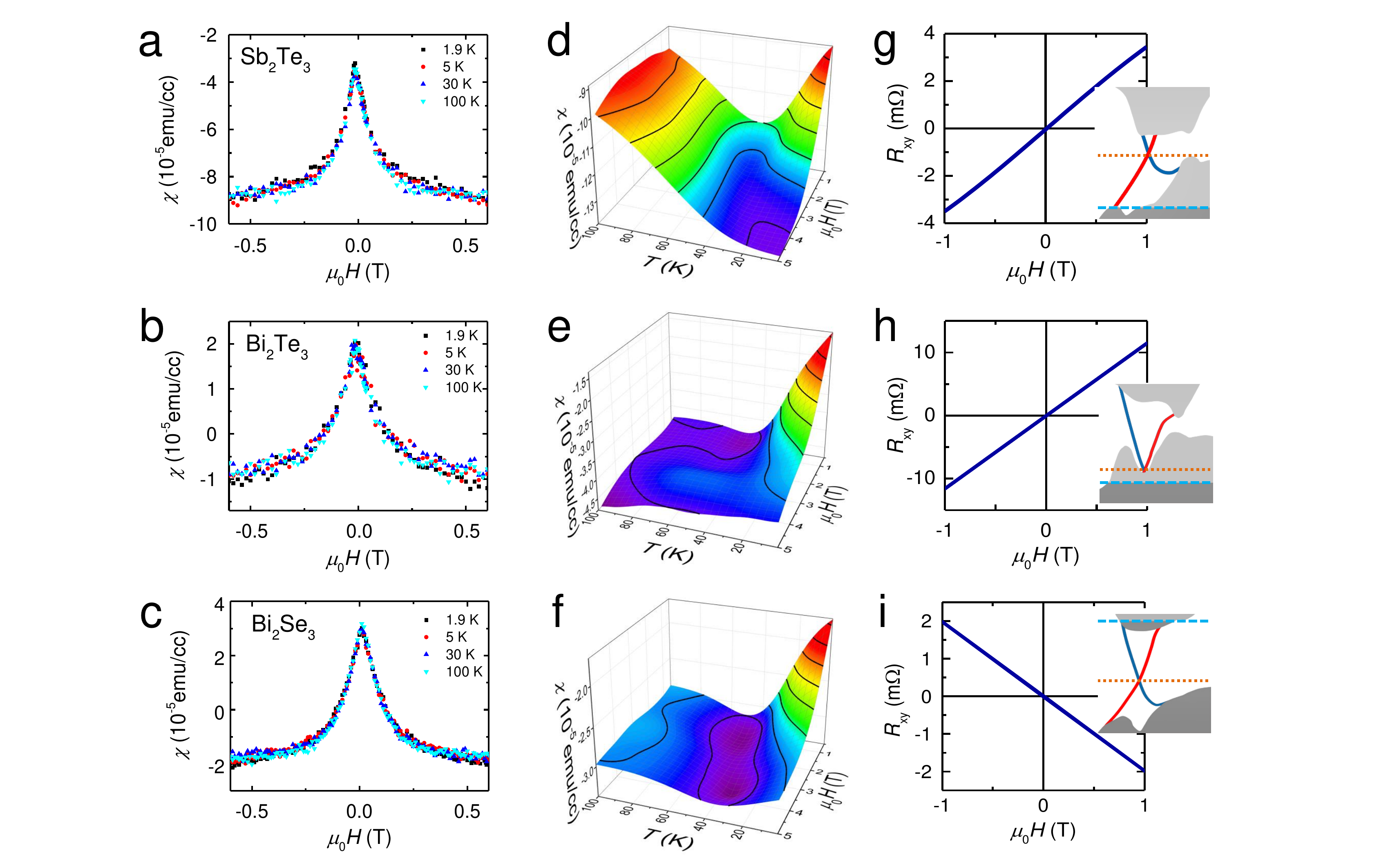}
\vfill\hfill Fig.~2 Zhao {\it et al.} \eject

\includegraphics[width=18cm]{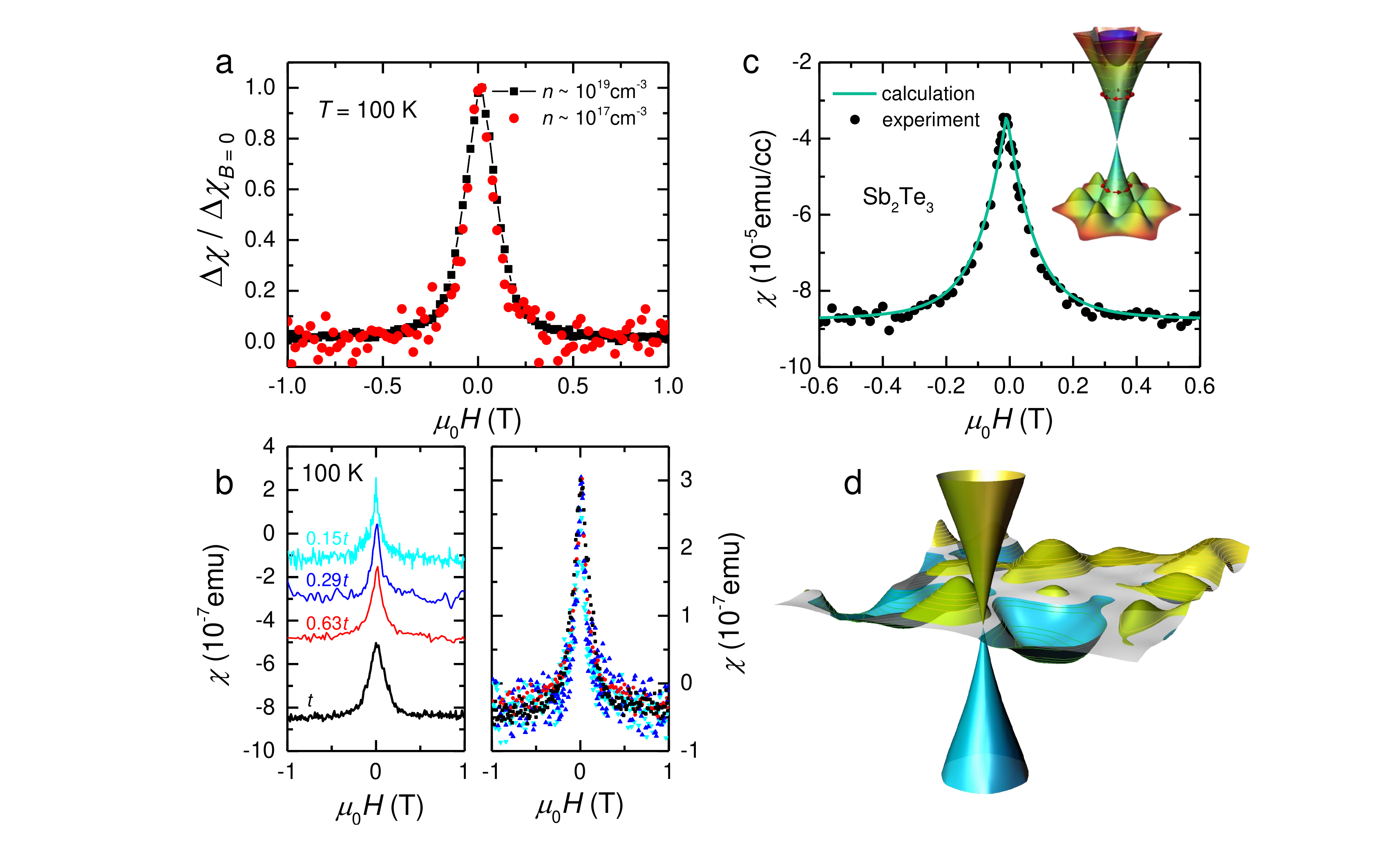}
\vfill\hfill Fig.~3 Zhao {\it et al.} \eject

\includegraphics[width=16cm]{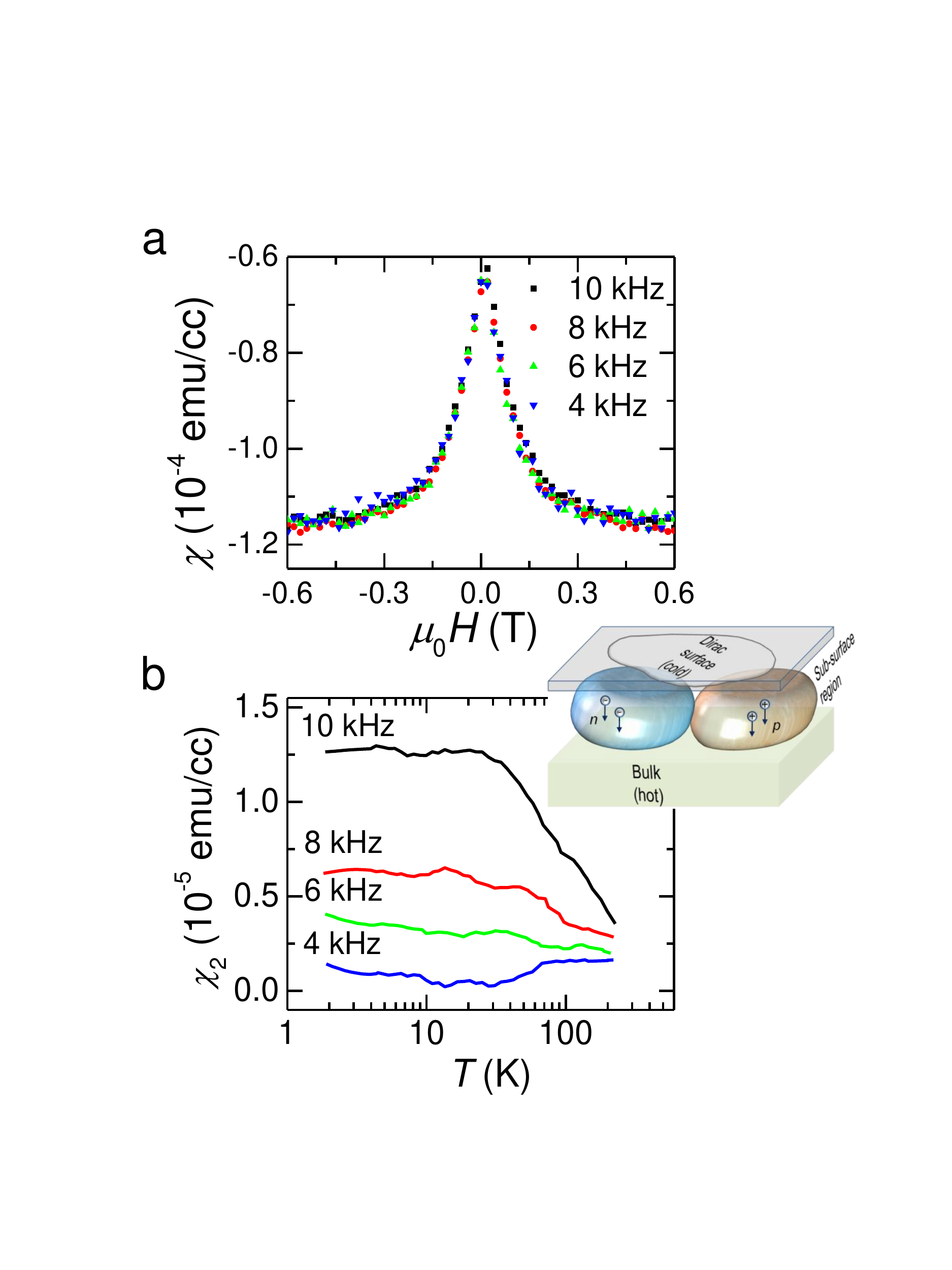}
\vfill\hfill Fig.~4 Zhao {\it et al.} \eject

\end{center}

\large\textbf{Supplemental information:}

The Supplementary Information is organized into experimental and theoretical parts, each organized into six and two subsections, respectively.
The experimental part presents calibration and background checks that verify the intrinsic origin of our 
findings and characterize our samples and our apparatus in more detail. The six subsections are:\\

(A) susceptibility calibration and background checks, (B) temperature dependence of the diamagnetic background, (C) observation of aging effects,
(D) determination of the \textit{g}-factor, (E) study of frequency dependence of susceptibility including  measurements obtained via strictly $dc$ probe, and
(F) susceptibility data from non-cleaving surfaces.\\

On the theory side, we include the calculation of singular Zeeman response from Dirac fermions, since to our knowledge this result has not previously appeared in the literature in this or other contexts. We provide a quantitative comparison to illustrate why Zeeman gap dominates Landau level gap in the case at hand.  Lastly, our proposal for Peltier cooling as a mechanism for maintaining singular response at elevated cryostat temperature is explained in some detail.

\section{Sample characterization, consistency checks and additional experiments}
\subsection{Susceptibility calibration and background checks}
\begin{figure}[h!]
\hspace{-15mm}
\includegraphics[width=1.0\linewidth]{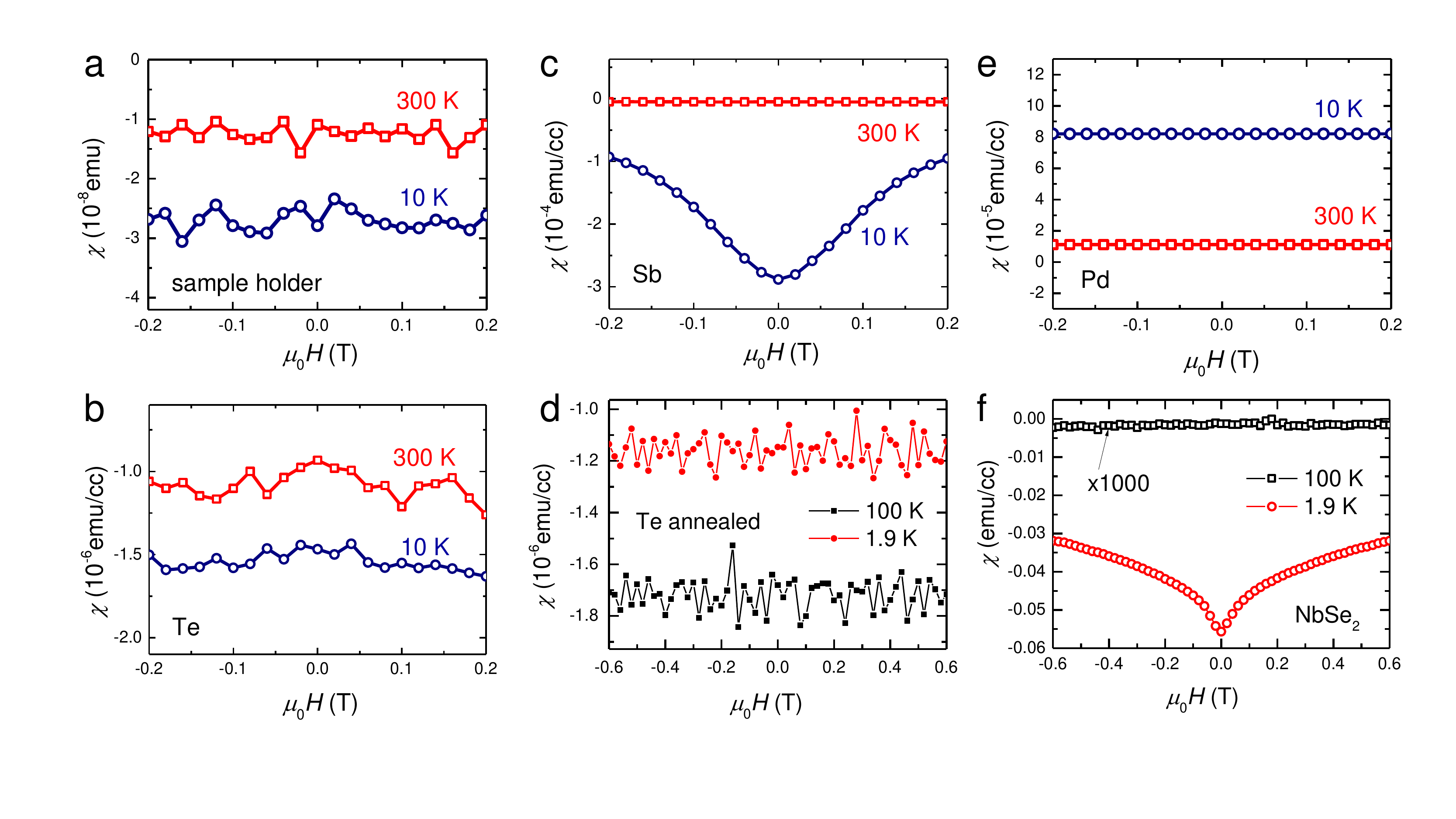}
\caption{\small
(a) Diamagnetic susceptibility of the sample holder used in the experiments is two orders of magnitude smaller than the typical signal from the sample signal. Susceptibility of precursor materials used in the crystal growth, Te and Sb, is depicted in panels (b,d) and (c), respectively. (d) Susceptibility of Te after annealing at $450\,^{\circ}{\rm C}$ in the growth furnace is also featureless. The well known large diamagnetism of Sb is enhanced at low temperatures (similar to Bi) while the diamagnetism of Te is only weakly temperature dependent. Both are in good agreement with the literature values. The signal from the paramagnetic Pd calibration sample is shown in (e). Susceptibility of the layered $2H-NbSe_2$ shows a well known behavior in the normal and superconducting states (see e.g. Ref. 31) and no paramagnetic cusp, see (f). These essentially featureless background/calibration checks are are in contrast with the cusplike paramagnetic field dependence at low fields consistently observed in the crystals of Sb$_2$Te$_3$, Bi$_2$Te$_3$, and Bi$_2$Se$_3$.
}
\protect
\label{fig:calibration}
\end{figure}
\normalsize
Our inductive lock-in measurement was calibrated using paramagnetic Pd sample. Additional measurements of other several materials were performed to establish uniqueness of singular low field response. We note that analyticity of the free energy as a function of magnetic field implies absence of singularities in the magnetic susceptibility and in many other physical quantities under common conditions. One certainly does not expect them to occur in the systems that do not host particle (spin) correlations. For example, a ferromagnetic material with small or null magnetic hysteresis will have a magnetization `jump' near zero field, and thus a singularity 
in differential magnetic susceptibility, $\chi$. Cusps can also occur in nonlinear susceptibility of frustrated spin systems (\textit{e.g.}, spin gasses, see Ref. [\onlinecite{spin-glass}]) where magnetic ions are present. Such magnetic correlation cusps are usually strongly temperature dependent below the correlation energy scale, which in spin glasses is typically well below room temperature. In our case, the glow discharge mass spectrometry analysis shows that no magnetic impurities are present to the level of $< 0.005~\textrm{ppm} ~\textrm{wt}$ level. Susceptibility data on Te and Sb in Fig.~\ref{fig:calibration} confirm this.

\subsection{Temperature dependence of diamagnetic susceptibility}
\begin{figure}[h!]
\includegraphics[width=\linewidth]{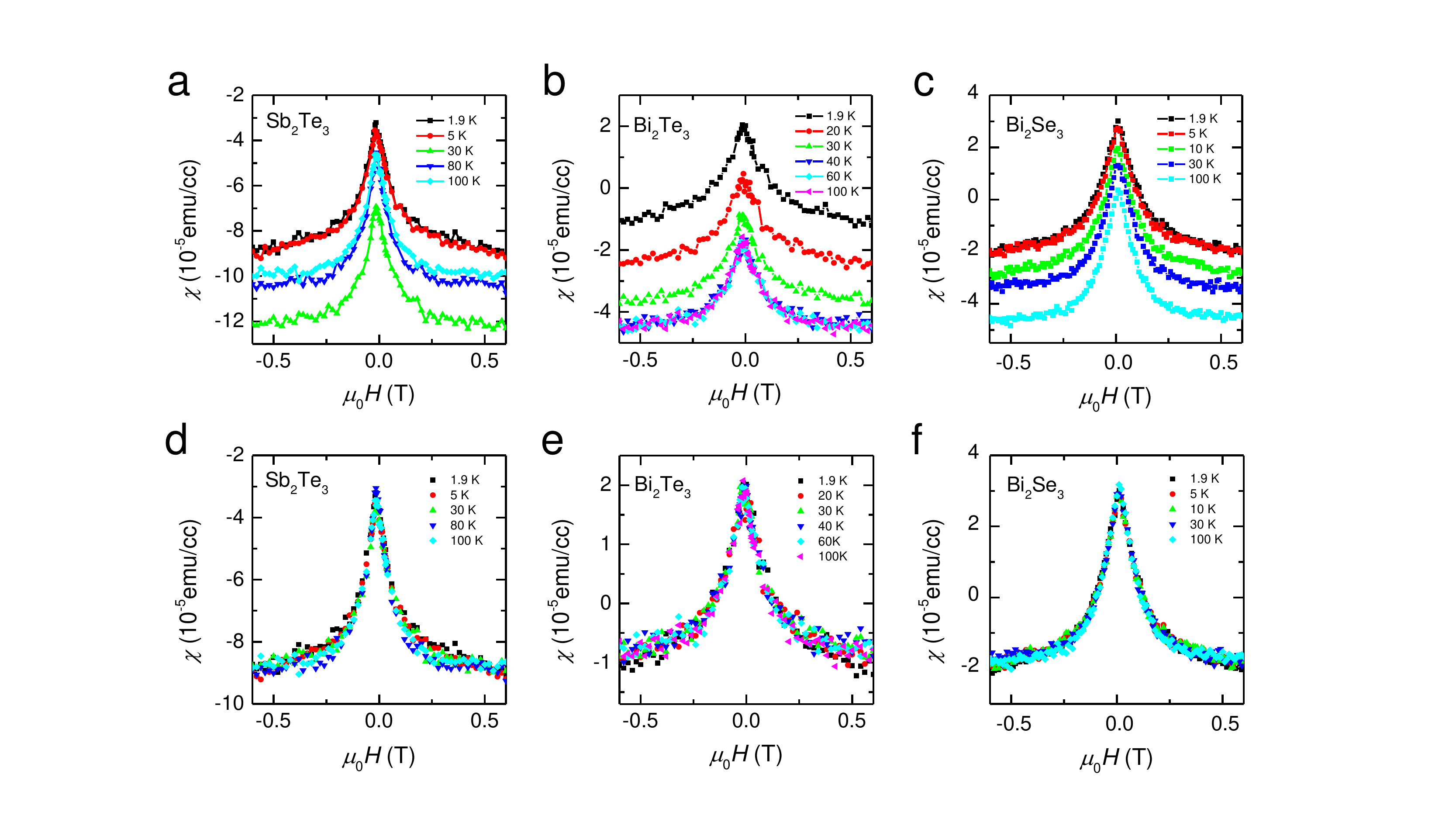}
\caption{\label{fig:Tdepdia}
\small The paramagnetic susceptibility cusp rides on a temperature dependent diamagnetic background, shown in (a) for Sb$_2$Te$_3$, in (b) for Bi$_2$Te$_3$, and in (c) for  Bi$_2$Se$_3$. The diamagnetism is largest for the Sb-based topological insulator, and smaller for the two Bi-based TIs.  Above $H \cong 0.5~ \textrm{T}$,  diamagnetic response depends on the details of band structure and on the position of chemical potential. However, the height and the slope of the cusp remain unchanged. To highlight the temperature robustness of the cusp, the data at higher temperatures in Fig.~1a, 1b, and 1c were shifted in Figs. 1-4 of the main text, as shown in (d), (e) and (f) respectively, to coincide with the susceptibility at 1.9 K (see main text).}
\protect
\end{figure}
\normalsize
The diamagnetic background shown in Figure S2 is temperature dependent. The rate of the temperature dependence clearly correlates with the bulk gap of these three materials -- larger gap usually implies weaker temperature dependence. By contrast, the paramagnetic anomaly at $B=0$ appears temperature independent in all three materials.

\subsection{Aging effect}
In all instances where we have measured the same sample more than once we have documented a clear aging effect by which the magnitude of the paramagnetic anomaly decreases with time. This is broadly reminiscent of aging effect documented in ARPES, whereby the electronic structure near the surface reconstructs via "band bending" similar to 2D heterostructure devices where there is also formation of similarly anistotropic states near the bottom of the bulk band. 
\begin{figure}[h!]
\vspace{-10mm}
\includegraphics[width=1.0\linewidth]{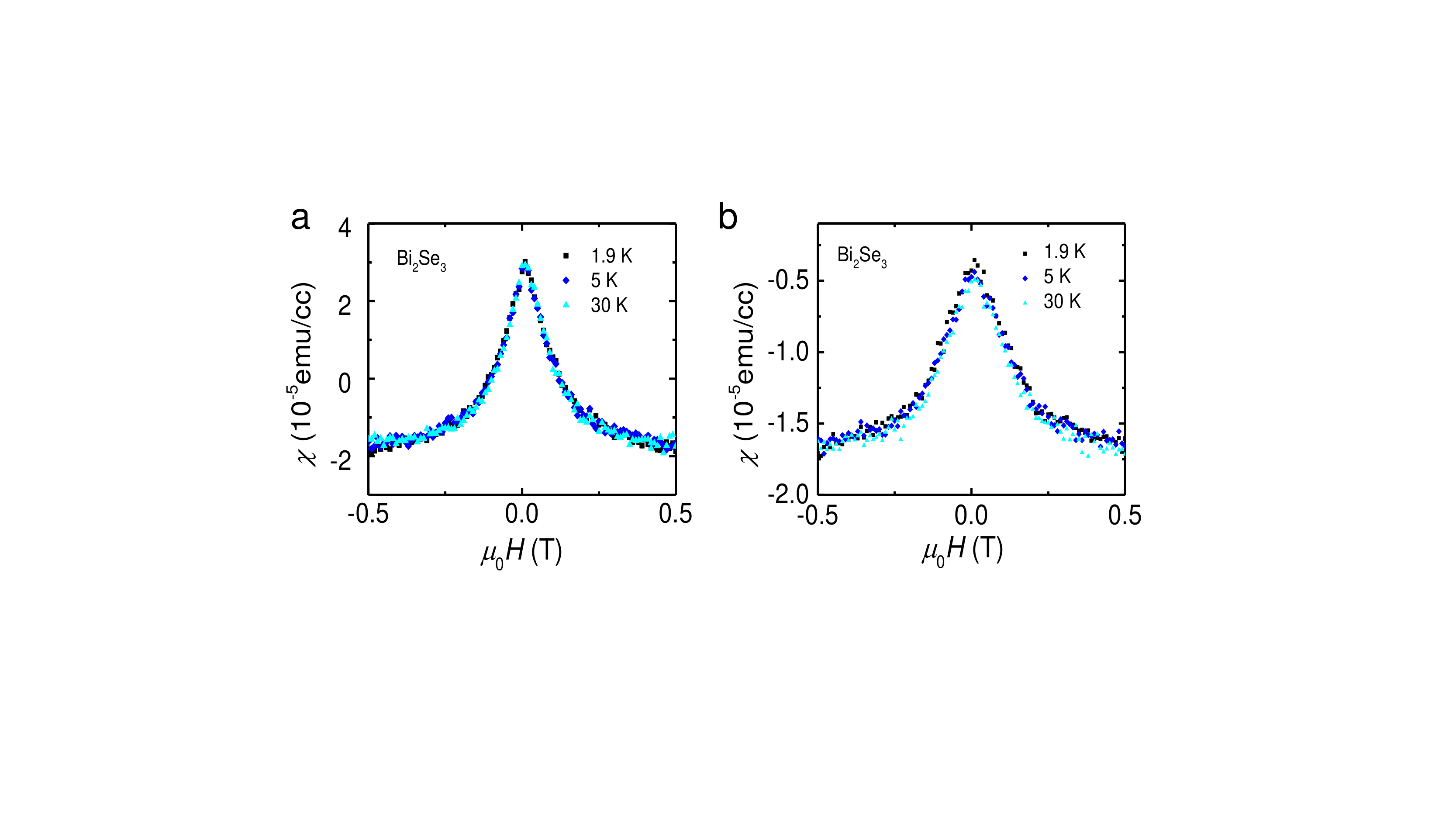}
\vspace{-20mm}
\caption{\label{fig:aging}
\small Susceptibility cusp for a Bi$_2$Se$_3$ crystal at several temperatures measured (a) an hour after the crystal growth and (b) two weeks later after the crystal was stored in flowing nitrogen. While the temperature robustness is intact, the overall cusp height has been reduced with time
likely due to surface reconstruction and the formation of two-dimensional electron gas (2DEG) associated with the band bending of the bulk states at the surfaces
\cite{surface-termination,Hofmann2DEG2012}.
}
\protect
\end{figure}
\normalsize
\subsection{Land\'{e} \textit{g}-factor from Shubnikov de Haas oscillations}
Topological insulators are expected to have strong Zeeman effects based on previously reported values of Land\'{e} \textit{g}-factor of about 60 \cite{Boebinger g-factor}.
 We have measured magneto-oscillations of the bulk conductivity and used it to deduce $g\approx 30$, fitting the oscillations with the Lifshitz-Kosevich equation \cite{Lifshitz} $\frac{\Delta \sigma_{xx}(T)}{\Delta \sigma_{xx}(0)}= \frac{\lambda(T)}{sinh \lambda(T)}$. Here
$\sigma_{xx}$ is the in-plane conductivity for magnetic field applied normal to the cleavage plane, $\lambda(T) = \frac{2 \pi k_BT}{\hbar eB}m_c$, and $m_c$ is the cyclotron mass. This is illustrated in Fig. \ref{fig:SdH} for Bi$_2$Te$_3$ crystal with carrier density $n \sim 10^{17}/cc$. Determination of the differences between the surface and bulk \textit{g}-values remains a challenge, as has been found in other studies.
\begin{figure}[h!]
\includegraphics[width=1\linewidth]{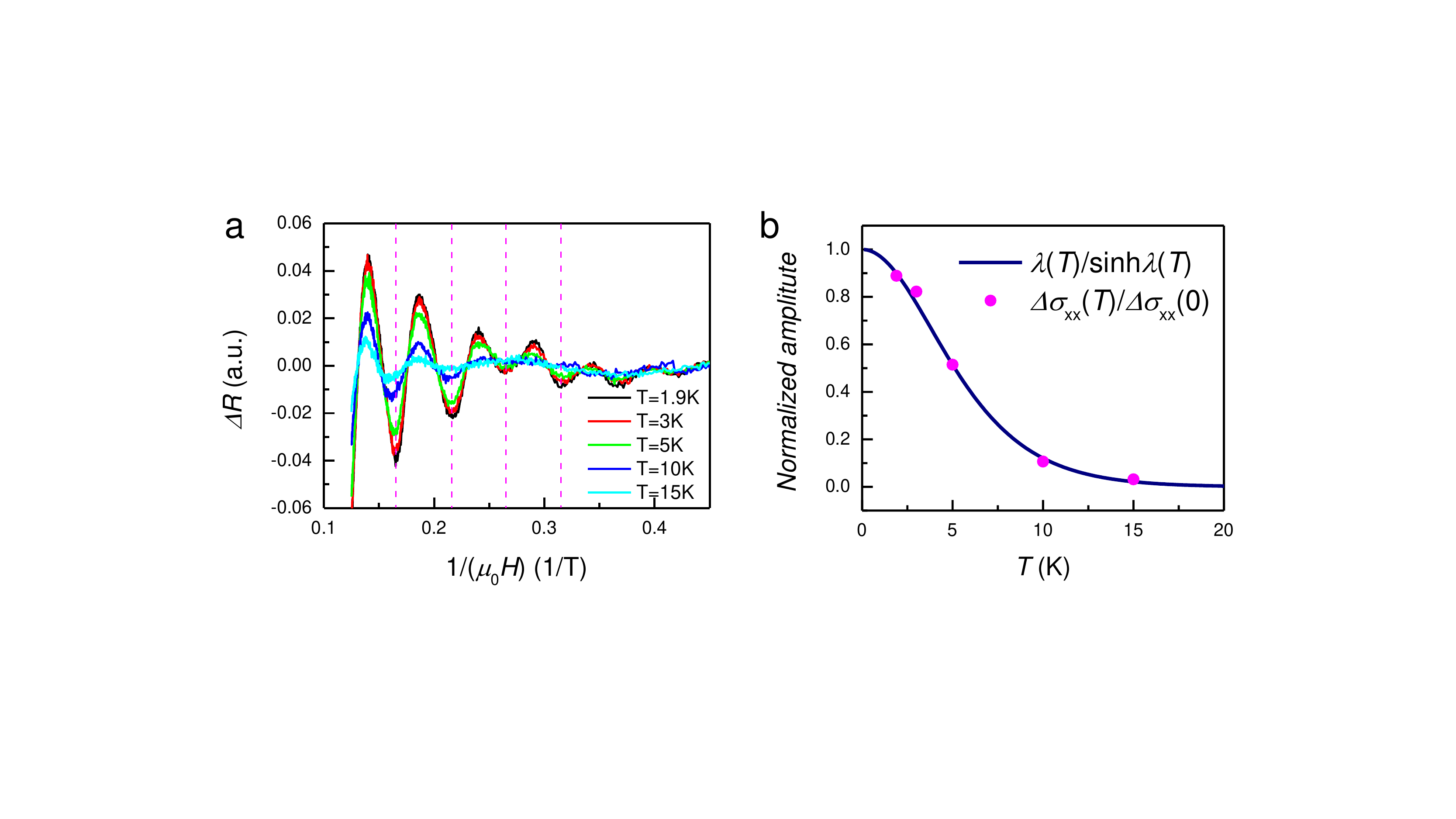}
\vspace{-40mm}
\caption{\label{fig:SdH}
\small (a) Shubnikov-de Haas (SdH) oscillations for a Bi$_2$Te$_3$ crystal with carrier density $n \sim 10^{17}/cc$; (b) The oscillations are fitted to Lifshitz-Kosevich formula using Monte Carlo technique. Below 8 Tesla field, the fit gives cyclotron mass $m_c = 0.0767 m_e$, where $m_e$ is bare electron mass. The obtained value of the \textit{g}-factor  is $g \sim \frac{2m_e}{m_c} \approx 30$, of the order of \textit{g}-factors $ \cong 60$ reported in other experiments \cite{Boebinger g-factor}.}
\protect
\end{figure}
\normalsize
\subsection{On the low frequency limit of the \emph{ac} magnetic susceptibility}
\begin{figure}[h!]
\vspace{-20mm}
\includegraphics[width=\linewidth]{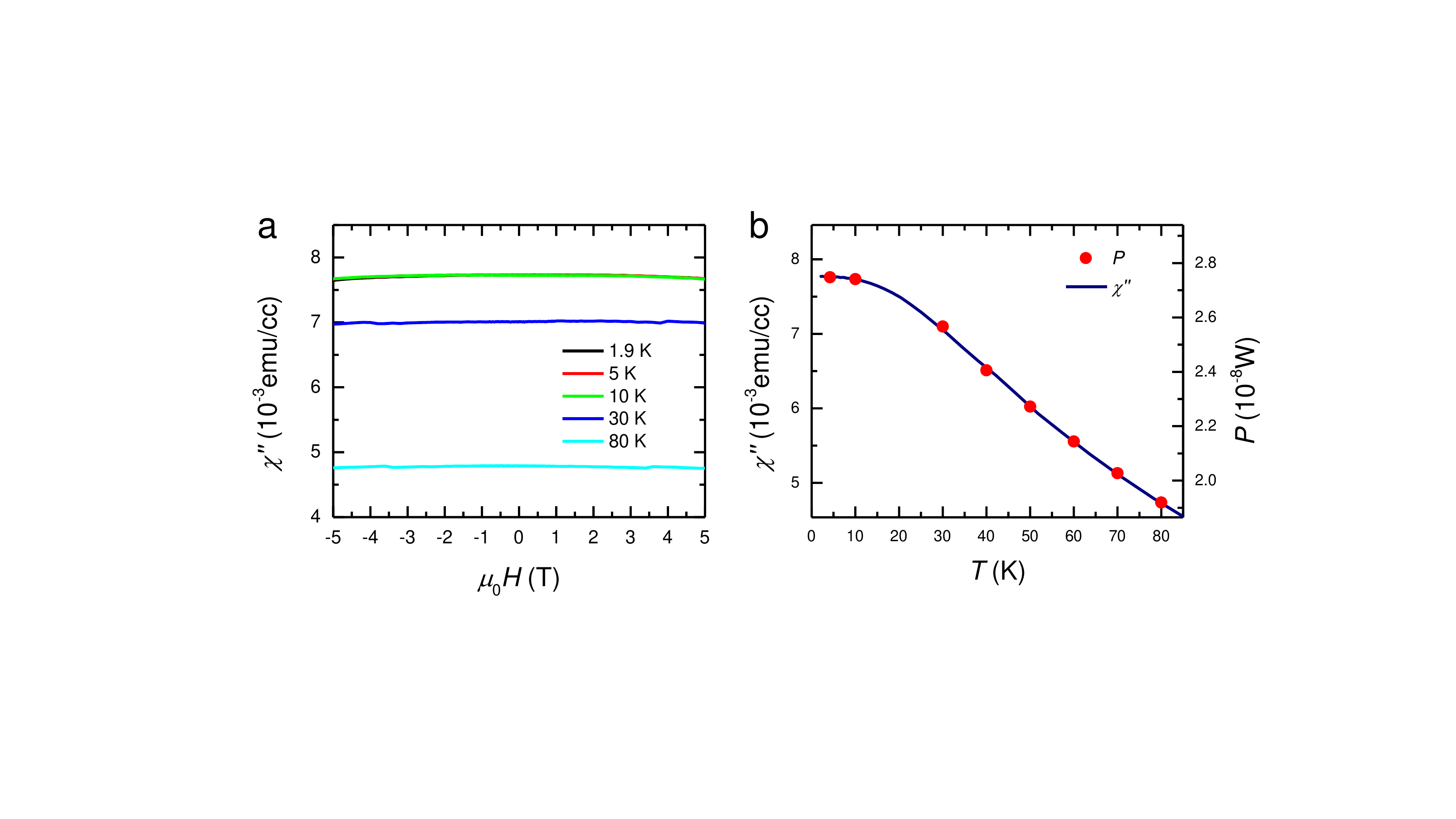}
\vspace{-25mm}
\caption{\label{fig:eddy}
\small (a) The out-of-phase susceptibility component of $\chi(\omega)$ is routinely recorded simultaneously with the in-phase component (here in Sb$_2$Te$_3$). It is purely dissipative and regular in the vicinity of $H = 0$; it does not display the cuspy behavior. (b) Assuming the standard eddy current mechanism is responsible for disspation, the out-of-phase component of $\chi(\omega)$ is proportional to conductivity, which is confirmed in our measurements: the observed value is consistent, up to geometric factors \emph{and} closely follows the temperature dependence 
in-plane resistivity $\rho_{xx}$, according to the standard formula for power $P = {\pi^2 (h_{ac}^2 d^2 f^2} / 2\rho_{xx} (T))$ dissipated during the \textit{ac} excitation cycles. Here  $f = \omega/2\pi $ and $d$ is the sample thickness.}
\protect
\end{figure}
\normalsize
Frequency dependent magnetic fields are screened by mobile charges on the scales set by the skin depth, estimated as $\sqrt{\frac{2}{\sigma \omega \mu}}$, where $\sigma$ is the sample's conductivity and $\omega$ is frequency. For our samples, and in the up to 10 kHz frequency range used, this value is on the order of a few millimeters, \textit{i.e.}, the field may be considered uniform inside the sample. Some residual frequency dependence of the diamagnetic background
may be due partly to frequency dependence of the skin depth. However, the cusp is frequency independent in the same frequency range. This is illustrated in Fig.~\ref{fig:freqsquid}b where all the data curves were shifted down to coincide with the 10 KHz data. Finite frequency magnetic response is necessarily complex, $\chi(\omega)=\chi_R(\omega)+i \chi_I(\omega)$, with an in-phase (real) and an out-of-phase (imaginary) components, $\chi_R$ and $\chi_I$, respectively. Finite $\chi_I$ signals dissipation (by eddy currents) and therefore must vanish, usually linearly in frequency. This is indeed the case as we checked explicitly. Its slope, $\chi_I(\omega)/\omega$, is quantitatively consistent (also as a function of temperature) with eddy current heating, see Fig.~\ref{fig:eddy}b. This dissipative component shows no sign of nonanalytic behavior as a function of magnetic field, see Fig.~\ref{fig:eddy}a.
\begin{figure}[h!]
\includegraphics[width=\linewidth]{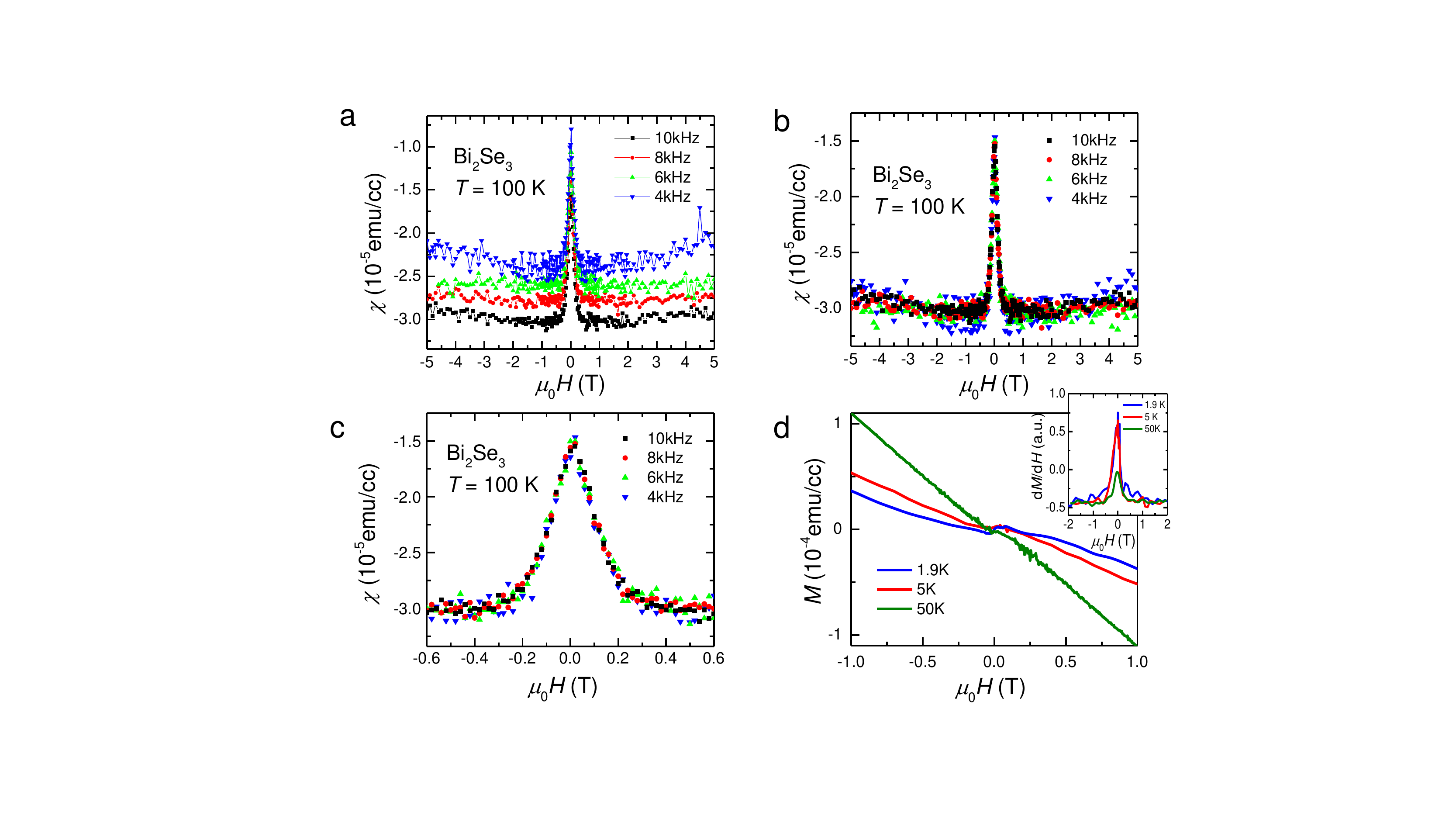}
\caption{\label{fig:freqsquid}
\small (a) Frequency dependence of the background and (b) independence of the singular contributions (see text for discussion) (c) Low-field blowup of data in (b). (d) \textit{dc} magnetization $M$ measured in the Superconducting Quantum Interference Device (SQUID) magnetometer shows clear nonlinearity near zero field. Numerical derivative $\frac{\textrm{d} M}{\textrm{d} H}$ of $M$ gives a `cusp', as shown in the inset. Consistent with our model in Section II the dc cusp singularity is rounded and diminished with decreasing temperature.
We note that taking numerical derivatives of magnetization in the vicinity of zero \textit{dc} magnetic field expectedly produces a spurious numerical noise. Hence, not surprisingly, data taken at a finite frequency, where the derivative is taking `in situ' using a small \textit{ac} oscillation gives a much more accurate record of magnetic susceptibility $\chi = \frac{\textrm{d} M}{\textrm{d} H}$ near $H \cong 0$. The derivative noise is partly controlled by the lock-in amplifier in the \textit{ac} detection circuit and it decreases at higher excitation frequencies.}
\end{figure}
\normalsize

The in-phase component is expected to provide a good estimate to the thermodynamic susceptibility, up to a small $\omega^2$ correction. This is indeed the case, see Fig. ~\ref{fig:freqsquid}a. Moreover, rather weak magnetoresistance 
implies negligible field dependence of the $\omega^2$ correction, \textit{i.e.} only a simple vertical offset is sufficient to compensate for the $\omega^2$ correction, see Fig. ~\ref{fig:freqsquid}b. The singular cusp, however, persists and is frequency independent, Fig.~\ref{fig:freqsquid}c.

It is important to note that the inductive technique we use is well known and commonly employed, \textit{e.g.} for measuring magneto-oscillations in metals, since 1940's. Thus far, nothing in the observed values or variation of $\chi(\omega)$ is surprising, except the low frequency cusp in $\chi_R$ vs. $B$ (we do, however, find second harmonic generation, see Section 2B, below). We note that the paramagnetic anomaly is also observed in \textit{dc} magnetization $M (H)$ measured using Superconducting Quantum Interference Device (SQUID) magnetometer, see Fig. ~\ref{fig:freqsquid}d. However, as can be seen from the figure, taking derivatives $\textrm{d}M/\textrm{d}H$ to obtain differential susceptibility near $H = 0$ is obviously numerically problematic, preventing direct quantitative comparison with the \emph{ac} traces.
\subsection{Singular spin response from noncleaving ``side" surfaces}
Noncleaving ``side" surfaces of TI crystals have received relatively little attention in part due to their poor quality, which makes them inaccessible to ARPES/STM.
In our experiments, samples rotated by ninety degrees clearly display a similar (albeit smaller) paramagnetic anomaly near $H = 0$ (Fig. S7c), which leads us to surmise that topological surface states are still present on side surfaces, and to our knowledge, this is the first observation of their response.
\begin{figure}[h!]
\includegraphics[width=1.0\linewidth]{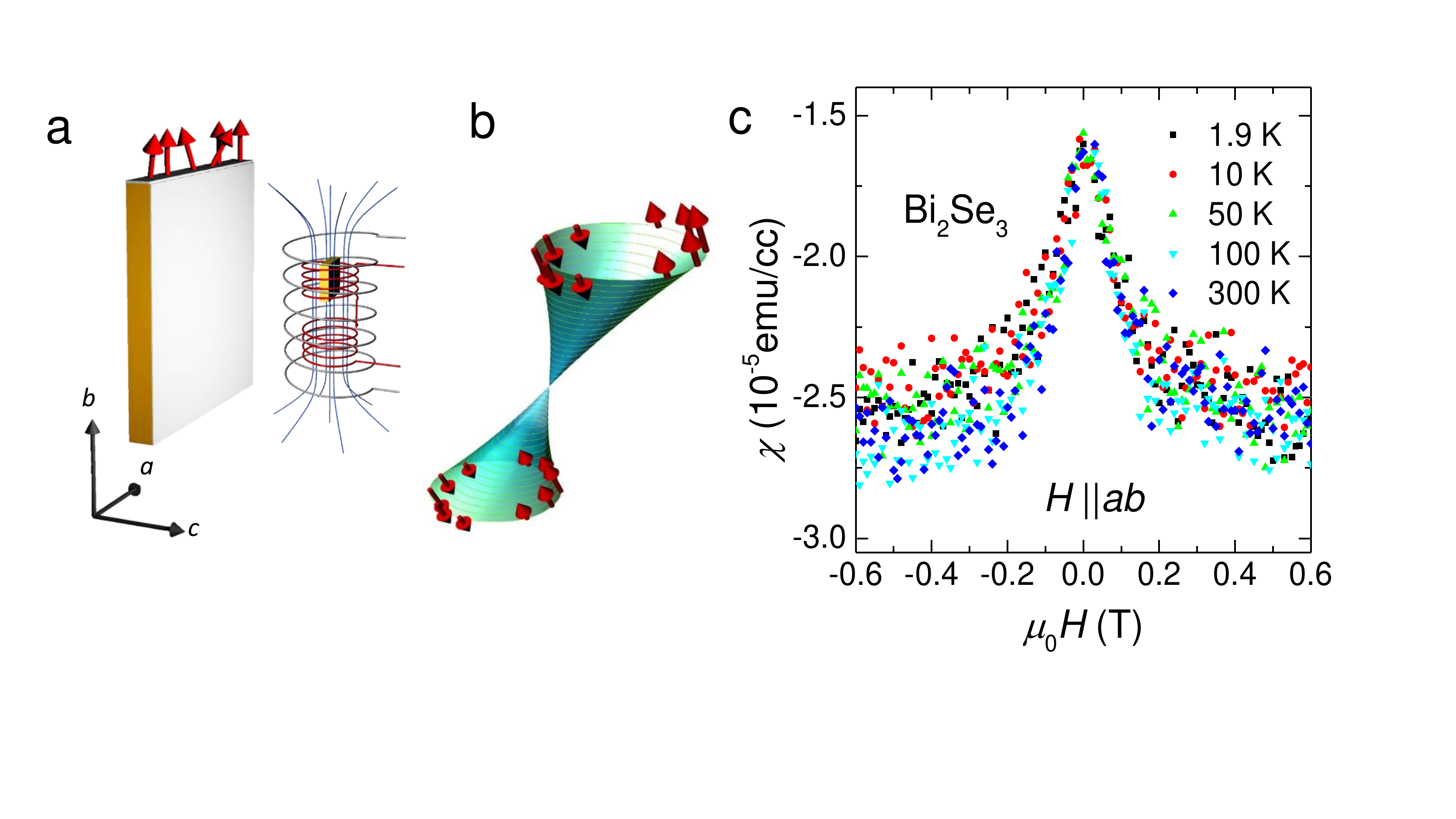}
\caption{\label{fig:sides}
(a) \textit{ac} measurement configuration with \textit{dc} magnetic field aligned normal to (00\={1}) ($H \parallel ab$-plane), for probing side surfaces of the same platelike crystal. (b) The constant energy contours on non-cleaving surfaces become elliptical with spin textures collapsed and tilted out of plane \cite{other-surfaces-2012}, as there is an intrinsic charge redistribution in the surface bands where different crystal faces connect. (c) The cusp for $H \parallel ab$, although smaller, also displays robustness against thermal rounding, although there may be some sign of rounding present.
}
\end{figure}
\section{Phenomenological theory}
\subsection{Singular paramagnetic response of Dirac spins}
Our experimental findings are notable not only in that singular response itself is detected but also that the same kind  of response (with a remarkably consistent large magnitude) persists across a broad swath of samples and three distinct families of topological materials. The physical phenomenon that underlies the data must transcend the unavoidable variations in screening, chemistry, preparation and such, making it particularly unlikely that the bulk of the samples makes any significant contribution to the low field singularity (see also Fig.~3b). Thus, our interpretation of the data aims squarely at the samples' surfaces, where interplay of universal physics of helical surfaces and large scale disorder appear to capture some of the more salient aspects of physics.

Surface states of ideal topological insulators are described by a band of helical Dirac fermions, minimally characterized by a simple Hamiltonian which assumes a particularly symmetric form for the (00\={1}) cleavage surface
\begin{equation}
H=\sum_{k,s,s'} (\hbar v_F \hat{n}\cdot \bf{k}\times \sigma_{ss'}-\mu \delta_{ss'}) c^\dagger_{k,s}c_{k,s'},
\label{eq:DHam}
\end{equation}
here $v_F$ is the Fermi velocity, $\mu$ is the chemical potential, $\sigma$'s are the Pauli spin matrices, $\hat{n}$ is the surface normal vector, and $c$ ($c^\dagger$) are the creation and annihilation operators. For this particular surface the helicity parameter, $|\hat{k}\times \sigma|$, is uniform in $\bf{k}$-space (which we take to be a disk with radius $\Lambda$) and the dispersion
is circularly symmetric near the Dirac point $({\bf k}=0)$.

For states sufficiently far away from the Dirac point the hybridization with bulk bands and various warping effects become pronounced (see, \textit{e.g.} Refs. \onlinecite{HexWarpFu2009,HexWarpKapitulnik2010,other surfaces 2012}), \textit{e.g.} hexagonal warping, while the electronic structure on non-cleaving surfaces is likely to be characterized by inhomogeneous helicity \cite{other surfaces 2012}. In this simple model these effects will enter implicitly as phenomenological parameters, such as the size of the \textbf{\={k}}-space unit cell $\Lambda$. In what follows we will not be including orbital quantization effects \cite{Franz PRB2012} -- we expect these to be unimportant on general grounds, namely owing to the presence of disorder and to large \textit{g}-factors in our samples, and they are indeed absent at low fields in our experiments.\begin{figure}[h]
\hspace{-20mm}
\includegraphics[width=0.75\linewidth]{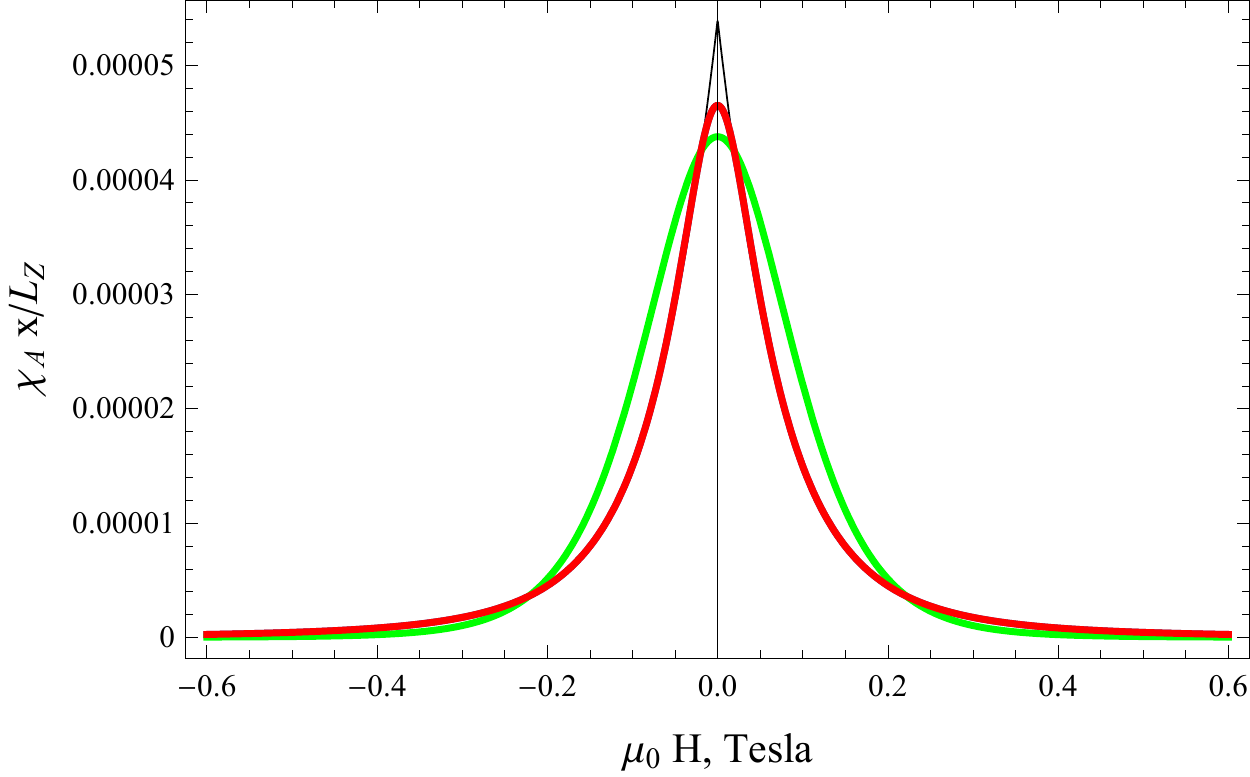}
\caption{\label{fig:theory}
\small Temperature dependence predicted in Eq. \ref{eq:chiT} is explored here by plotting traces at $T=0$ (black), $T=1K$ (red) and $T=10K$ (green and red, see below). We compare the first plot to choose experimental parameters,\textit{i.e.} $g=60$, $\Lambda=5\cdot 10^8 m^{-1}$, $v_F=2000 m/s$, $x=0.18\cdot 10^{-3}$, also $L_z\approx1~mm$ is the measured thickness of these samples. These parameters are sample dependent -- they can vary with sample growth and preparation, as well as with the level of surface reconstruction and/or ``aging" \cite{Hofmann2DEG2012}.  The value of $\Lambda$, the characteristic size of the momentum space of the surface states,  is a few percent of the corresponding bulk quantity, its microscopic meaning remains to be established. Here, we pick it to be some fraction of the bulk Brillouin zone. Lastly, the Fermi velocity we use is consistent with that obtained from a low energy probe, see Ref.~\onlinecite{Wolos2012}. If these parameters are rescaled upon raising temperature (from 1K to 10K), $\Lambda\to\Lambda,\ v_F\to 10 v_F,\ x\to x/10 ,\ g\to 10 g$,  $\chi$ is invariant (red trace). Otherwise, in the green trace, we explore whether approximate invariance may be maintained under less fine-tuned rescaling, \textit{e.g.} with $\Lambda\to\Lambda/2,\ v_F\to2 v_F,\ x\to 1.5 x,\ g\to 3 g$. This discussion is not intended as analysis of the experiments, but rather to explore potential thermal effects theoretically.
}
\end{figure}
\normalsize
The Zeeman coupling is introduced via
\begin{equation}
H_Z=(g \mu_B B/2) \sum_{k,s,s'} \hat{n}\cdot \sigma_{s,s'}c^\dagger_{k,s}c_{k,s'}.
\end{equation}
It opens a gap near the Dirac point between conduction and valence bands, $s=\pm 1$, respectively
\be
\epsilon_{k,s}=s \sqrt{(g \mu_B B)^2+(\hbar v_F k)^2}\equiv s \epsilon_k.
\ee

Quantitatively, we may gauge the relative importance of the orbital \textit{vs}. Zeeman contribution by comparing gaps (Landau \textit{vs}. Zeeman).
While Landau quantization is expected to dominate at sufficiently low fields, the relevant field scale
\be
\hbar v_F \sqrt{\frac{e B}{\hbar}}> g \mu_B B\to B<\frac{e\ v_F^2 \hbar}{(g \mu_B)^2}\lesssim 10^{-4} {\rm T}
\ee
is much lower than our experimental resolution (we have used large $g\approx 60$ and small $v_F\approx 2000 ~\textrm{m/s}$, which is an appropriate ballpark for our samples in the $\mu = 0$ patch regions, see above for determination of $g$ and below for fit and discussion of $v_F$).

The exact expression for the areal (sheet) susceptibility of a single 2D Dirac state $\chi_A=\partial M/\partial H=-\mu_0 \partial^2 F/\partial B^2$ can be obtained, where
\begin{align}
&M=-\frac{\partial}{\partial B} (E- k_B T S)
=-(g \mu_B)^2 B \sum_{s=\pm} \int \frac{d^2 k}{(2\pi)^2}\frac{s}{\epsilon_k}\tanh \frac{\beta}{2} (s \epsilon_k -\mu)\\
&=-\frac{ (g \mu_B)^2  B}{\pi \beta \hbar^2 v_F^2} \sum_{s=\pm} \log \cosh \frac{\beta}{2}(s 
{y}-\mu)\Big|_{y=\epsilon_0}^{y=\epsilon_\Lambda}
\xrightarrow[\mu=0]{\beta\to\infty}
\frac{(g \mu_B)^2  B}{\pi \hbar^2 v_F^2} (\sqrt{(g \mu_B B)^2+(\hbar v_F \Lambda)^2}-g \mu_B |B|)
\label{eq:chiT}
\end{align}
Weak magnetic field acts perturbatively as long as $\mu\neq0$ in that spin-orbit locked electrons at the Fermi level polarize only slightly (far less than in the spin degenerate Fermi gas), hence $\chi(B)$ is analytic as $B\to0$. For very large $B$ the response follows  $\sim 1/|B|^3$ (as typical of Van-Vleck paramagnetism \cite{VanVleck}). The transition between these two behaviors takes place at $B_C=\pm \mu/(g \mu_B)$ via a jump singularity in $\chi_A$. The singular response at $\mu=0$ descends from these singularities.

At $k_B T=\mu=0$ the susceptibility reduces to
\begin{equation}
\chi_A(B)=\frac{\mu_0(g \mu_B)^2 (-2 g \mu_B|B| \sqrt{(g \mu_B B)^2+(\hbar v_F \Lambda)^2}+2 (g \mu_B B)^2+(\hbar v_F \Lambda)^2)}{4\pi^2 \hbar ^2 v_F^2 \sqrt{(g \mu_B B)^2+(\hbar v_F \Lambda)^2}}
\label{eq:chi0}
\end{equation}
which has the form of susceptibility data shown in Figs. 1-4,\ref{fig:freqsquid},\ref{fig:Tdepdia},\ref{fig:aging},\ref{fig:sides}.
In particular, the hallmark of Dirac physics is the universality of the slope of the $\sim |B|$ term, which only depends on the \textit{g}-factor and the Fermi velocity and not on the size of Brillouin zone, while the maximum of susceptibility $\chi_A(0)$ at $B = 0$ depends on the details of warping and hybridization with the bulk through $\Lambda$.

We now turn to a more quantitative exploration of this phenomenology, to establish the existence of reasonable choice of parameters that can reproduce the experimental results. As already discussed in the main text, we postulate the existence of regions with $\mu\cong0$ that are sufficiently large so that this (nominally translationally invariant) theory applies and the net response can be approximated as arithmetic average over various contributions from the bulk and surfaces, $\chi(B)=\chi_0+\chi_A(B) ~x/L_z$. Most of the surface is significantly detuned from the Dirac point and only contributes to the (non-singular) background, as discussed above. Thus, we introduce one additional parameter, the corresponding surface fraction, $x<1$, of $\mu\cong0$ regions. Prevalence of singular response then points to ubiquity of such regions; however to actually locate these regions and elucidate the physics responsible for their formation is outside the scope of our Dirac phenomenology and will require further theoretical and experimental work.

The apparent susceptibility, without all non-singular background contributions, is $\delta \chi(B)=\chi_A(B)~x/L_z$, where $L_z$ is sample's thickness.

\subsection{Efficient cooling of Dirac fermions}
\begin{figure}[h]
\vspace{-10mm}
\includegraphics[width=.8\linewidth]{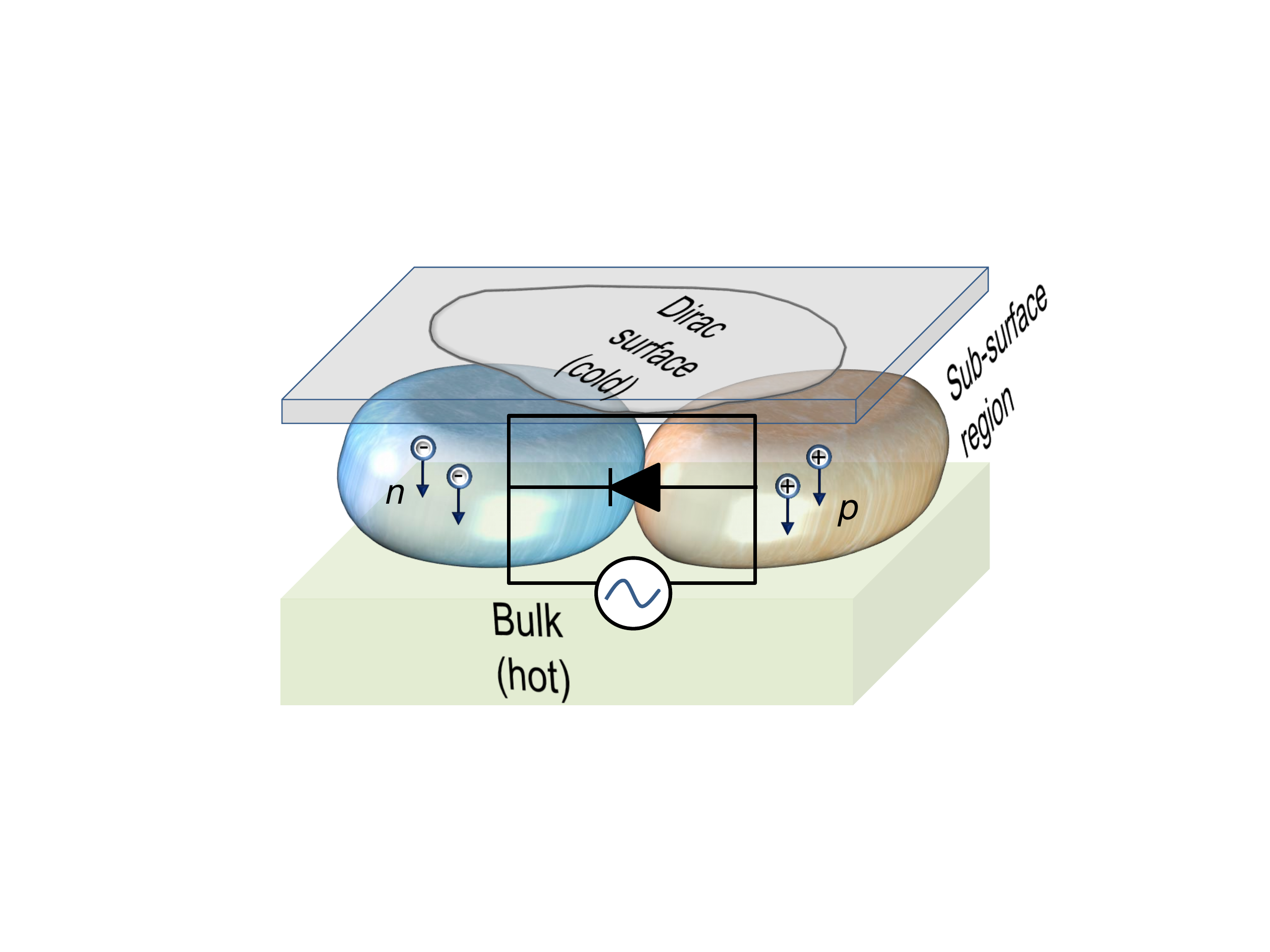}
\vspace{-25mm}
\caption{\label{fig:cooler}
\small The basic idea is that a local fluctuation in doping, \textit{e.g.} of the kind that may create a $\mu\cong0$ patch at the surface (see above), will also create adjacent $p$ and $n$ doped regions. With the $ac$ excitation field oriented along the $c$-axis we expect induced currents primarily flowing in the $ab$ plane.  For the experimental parameters, such as material's conductivity and probe's  frequency range, the dissipative out-of-phase (eddy current) component dominates the in-phase component by about 3 orders of magnitude.  These finite frequency currents are expected to exhibit inhomogeneities induced by variations in electronic structure. In particular, it is natural to find current loops traversing through the bulk, the surface and the said $p-n$ regions.  Under favorable conditions such current loops will act as mesoscopic Peltier coolers.  These favorable conditions include relatively low local resistance (and smooth disorder) along the current path that helps focus the current flow through thermoelectrically asymmetric region (as drawn in Fig. S9) and also direct contact between $p$ and $n$ regions that result in formation of a depletion layer and rectification path (diode shunt in the figure).}
\end{figure}
\normalsize

Room temperature stability of the singular cusp is difficult to achieve within the confines of theory outlined in the previous section. In fact, the challenge is considerably more general -- the thermal energy at $T=300~\textrm{K}$ is an appreciable fraction of the bulk gap of these narrowband materials.
This results in significant
temperature dependence of `background' diamagnetic susceptibility, conductivity and other transport properties. We surmise that (Dirac) surface states responsible for the singularity are maintained at a different (significantly colder) temperature than the rest of the sample. We note a very recent study \cite{thermoelectric-GedikPRL2012} where uniquely slow, power-law in time, energy relaxation out of excited surface states in TIs was detected and attributed to acoustic-phonon-dominated coupling to the bulk. Such a weak coupling is a prerequisite for our proposal.

We now sketch out a simple and plausible, albeit speculative, scenario by which the $ac$ nature of the probe itself, and, more specifically, eddy currents as observed in out-of-phase component of $ac$ susceptibility, combined with subsurface disorder which essentially provides for proximate $p$ and $n$ regions, act to cool the surface electronic excitations responsible for the cusp well below the sample's bulk temperature. While systematic studies of surface mesoscopics are needed to flesh out and test the various aspects of this scenario, it is certain that some kind of a (non-equilibrium) cooling process is operative based on the energy scales argument above and but also on our additional experimental observations of slow equilibration, dissipative eddy current response and sizeable harmonic generation.
The picture below is the simplest example of how the combination of known good thermoelectric properties of these materials, disorder morphology, simple semiconductor facts \emph{and} $ac$ nature of our probe may produce the sought after cooling behavior.
\begin{figure}
\includegraphics[width=\linewidth]{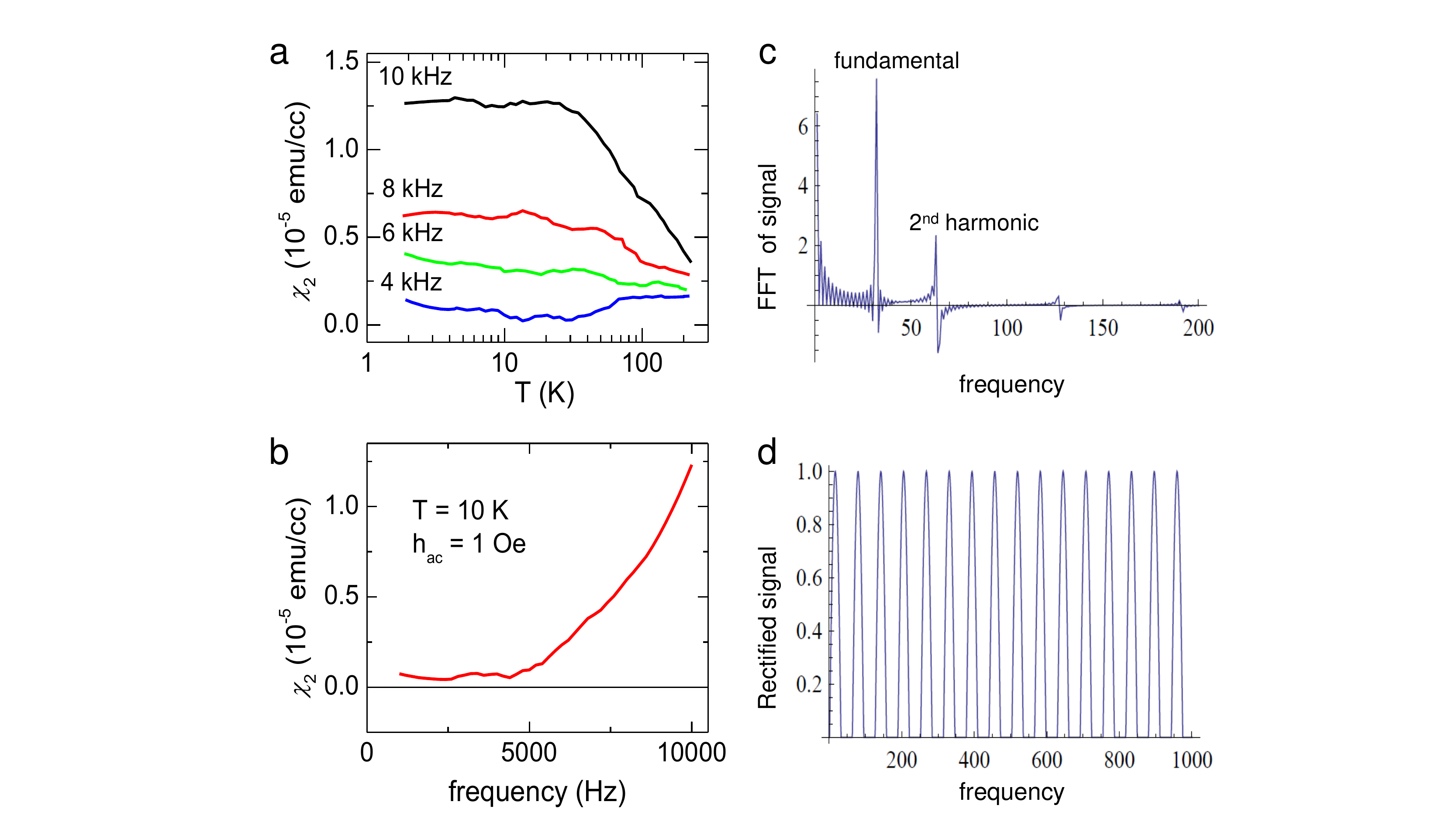}
\caption{\label{fig:coolerdetails}
\small (a) The nonlinearity of the surface-bulk connection is witnessed by the observed $2^{nd}$ harmonic $\chi_2$. (b) As expected, $\chi_2$ is quadratic in frequency. It is consistent with the existence of ``rectifying" paths in the putative thermoelectric cooling elements (Fig. S9) required for the cooling of small fraction of sample's surface and thus suppressing thermalization of Dirac surfaces with the bulk. The generation of $2^{nd}$ harmonic (c) is associated with (d) the rectified signal, see discussion in the text.
}
\protect
\end{figure}
\normalsize

Unlike conventional (macroscopic) Peltier coolers, where \textit{p} and \textit{n} regions are well separated and no depletion layer forms, here we need direct contact if this `device' is to operate on alternating currents source provided by eddy response (conventional Peltier cooler require DC power source). Such direct contact will form an effective rectifying element -- a diode shunt -- which will redirect the electric current away from the Peltier cooling path during the ``wrong" half of the cycle (when it would otherwise act as a heater). The detected $2^{nd}$ harmonic generation (shown,\textit{ e.g.,} for Sb$_2$Te$_3$ in panels (a) and (b) of Fig.~S10, and in Fig. 4) is consistent with this scenario. Generation of second harmonic through rectification is illustrated by a simple calculation that shows that (Fig.~10c) the presence of the $2^{nd}$ harmonic can originate from signal rectification (Fig.~S10d). Dissipative out-of-phase (eddy current) response, shown for Sb$_2$Te$_3$ in Fig. S5, is fully consistent with both the values and the temperature dependence of the in-plane resistivity $\rho_{xx}$, giving a temperature dependent power dissipation. More elaborate configurations of \textit{p} and \textit{n} regions capable of simultaneous rectification and cooling may be imagined, of course. However, one particularly appealing aspect of this simplest ``two-blob" device is the internally constrained match of polarity that ensures simultaneously correct signs of heat transfer and rectification.

Maintaining the singular cusp response implies the need to keep relevant electrons at very low cryogenic temperature. This requires a highly effective cooling `device', much more than what's presently available on macroscales. Cooling efficiency can be  enhanced by a number of other effects, such as mesoscopic self-compatibility \cite{Thompson_cooler} and nanoconstrictions \cite{nano-constrictions}. However, the observed unusually strong harmonic generation suggests that larger values of thermoelectric parameters may originate from strong frequency dependence of the transport coefficients under geometric confinement. For phonons such resonances are natural and well known \cite{frequency-dependent-thermal}. Recent considerations of the spin Seebeck effect show that in contrast with bulk Seebeck effect, the figure of merit of nanoscale thermal-spin conversion can be infinite, leading to the ideal Carnot efficiency \cite{spin-rectifier} (in the nonlinear spin Seebeck transport regime the system acts as a nanoscale thermal spin rectifier). Finally we note that thermopower is strongly affected by the spin-orbit coupling \cite{nano-constrictions}, with asymmetry provided by the non-degenerate spin channels, leading to much larger cooling enhancements on mesoscale.



\end{document}